\documentclass[preprint,12pt]{elsarticle}

\usepackage{amssymb}
\usepackage[utf8]{inputenc}
\usepackage{soul}
\usepackage{color}
\usepackage{graphicx}
\usepackage{amsmath}
\usepackage[version=4]{mhchem}
\usepackage{siunitx}
\usepackage{longtable,tabularx}
\usepackage{caption}
\usepackage{float}
\usepackage[nameinlink,capitalise,noabbrev]{cleveref}
\usepackage[dvipsnames]{xcolor}
\usepackage{algorithm}
\usepackage{algorithmic}
\graphicspath{{Pictures/}}
\usepackage{subcaption}
\usepackage{multirow}
\usepackage{booktabs}
\usepackage{xcolor}
\usepackage{geometry}
\usepackage{array}
\usepackage{textcomp}
\usepackage{ulem}
\usepackage{placeins}
\usepackage{array}
\newcolumntype{P}[1]{>{\centering\arraybackslash}p{#1}}
\newcolumntype{M}[1]{>{\centering\arraybackslash}m{#1}}

\begin{document}

\begin{frontmatter}

\title{Methods for Multi-objective Optimization PID Controller for quadrotor UAVs} 

\author[inst1]{Andrea Vaiuso\corref{cor1}}
\ead{vaiu@zhaw.ch}
\fntext[label1]{Research Associate}
\cortext[cor1]{Corresponding Author}

\author[inst1]{Gabriele Immordino\fnref{label2}}
\fntext[label2]{Post-doc}

\author[inst1,inst3]{Ludovica Onofri\fnref{label5}}
\fntext[label5]{PhD Student}

\author[inst3]{Giuliano Coppotelli\fnref{label4}}
\fntext[label4]{Associate Professor, AIAA Associate Fellow}

\author[inst1]{Marcello Righi\fnref{label3}}
\fntext[label3]{Professor, AIAA Senior Member, Lecturer at Federal Institute of Technology Zurich ETHZ}

\affiliation[inst1]{organization={School of Engineering, Zurich University of Applied Sciences ZHAW},
            city={Winterthur},
            country={Switzerland}}

\affiliation[inst2]{organization={Faculty of Engineering and Physical Sciences, University of Southampton},
            city={Southampton},
            country={United Kingdom}}

\affiliation[inst3]{organization={Department of Mechanical and Aerospace Engineering,"La Sapienza" University of Rome },
            country={Italy}}

\begin{abstract}

Integrating unmanned aerial vehicles into daily use requires controllers that ensure stable flight, efficient energy use, and reduced noise. Proportional integral derivative controllers remain standard but are highly sensitive to gain selection, with manual tuning often yielding suboptimal trade-offs. This paper studies different optimization techniques for the automated tuning of quadrotor proportional integral derivative gains under a unified simulation that couples a blade element momentum based aerodynamic model with a fast deep neural network surrogate, six degrees of freedom rigid body dynamics, turbulence, and a data driven acoustic surrogate model that predicts third octave spectra and propagates them to ground receivers. We compare three families of gradient-free optimizers: metaheuristics, Bayesian optimization, and deep reinforcement learning. Candidate controllers are evaluated using a composite cost function that incorporates multiple metrics, such as noise footprint and power consumption, simultaneously. Experiments are run in a complete set of missions, rather than studying a single response, spanning wide, unconstrained searches and bounded refinements, using a manually tuned Ziegler–Nichols baseline for reference. Metaheuristics improve performance consistently, with Grey Wolf Optimization producing optimal results. Bayesian optimization is sample efficient but carries higher per iteration overhead and depends on the design domain. The reinforcement learning agents do not surpass the baseline in the current setup, suggesting the problem formulation requires further refinement. On unseen missions the best tuned controller maintains accurate tracking while reducing oscillations, power demand, and acoustic emissions. These results show that noise aware proportional integral derivative tuning through black box search can deliver quieter and more efficient flight without hardware changes.

\end{abstract}

\end{frontmatter}

\section{Introduction}


Unmanned Aerial Vehicles (UAVs) are increasingly recognized as versatile, multifunctional, and cost-effective tools, with applications in transportation, communication, agriculture, disaster response, and environmental monitoring~\cite{floreano2015science}. Their integration into civil and commercial sectors continues to expand, offering sustainable alternatives that can reduce greenhouse gas emissions and mitigate other environmental impacts~\cite{goodchild2018delivery}. However, their widespread adoption in urban and suburban environments faces challenges related to efficiency, reliability, and public acceptance related to acoustic disturbance \cite{watkins2020ten}, that affects both humans and wildlife \cite{raya2017small}. These limitations highlight the need for advanced UAV design techniques and controller tuning strategies \cite{maaruf2022survey}, that can simultaneously account for different, and often conflicting, requirements. As a result, operating UAVs in densely populated environments poses an inherently multi-objective problem where competing performance criteria must be optimized and balanced simultaneously.

A large body of research has focused on hardware-oriented solutions to address the problem of noise pollution, including aerodynamic refinements, propeller geometry optimization, and the use of noise-absorbing materials \cite{uragun2014noise,gur2009design,trinh2025designs, mane2024comprehensive}. Although effective, such modifications often involve significant costs and may compromise aerodynamic performance.  Therefore, control strategies are often preferred as a more flexible and less invasive alternative. Proportional–Integral–Derivative (PID) controllers remain the most widely adopted solution for UAV stabilization due to their simplicity, robustness, and ease of implementation \cite{abdelhay2019modeling, sahrir2022modelling, lopez2023pid}. 

Optimization-based control approaches have been extensively studied in literature, as a means to enhance the performance of existing UAV platforms through systematic controller tuning rather than structural redesign \cite{george2022optimal}.
Classical PID tuning approaches, such as the Ziegler–Nichols technique \cite{meshram2012tuning} and other rule-based heuristics like Cohen-Coon and Gain and Phase methods \cite{foley2005comparison}, provide simple but straightforward procedures to determine gain values. However, they typically rely on empirical tests, are sensitive to model uncertainties, and tend to produce aggressive control actions that may lead to overshoot, oscillations, or inefficient energy use especially in nonlinear dynamic models \cite{joseph2022metaheuristic}. Lastly, these methods do not explicitly account for multi-objective trade-offs, including noise reduction and power efficiency.

One of the most widely explored methods for PID controls tuning is optimization based on metaheuristics \cite{abushawish2020pid, joseph2022metaheuristic}. There exists a large variety of metaheuristic algorithms, ranging from evolutionary approaches to swarm intelligence and physics-inspired methods, each with its own strengths and trade-offs. Among them, Particle Swarm Optimization (PSO) \cite{Kennedy2002ParticleSO} remains nowadays the most widely adopted technique in PID controller tuning \cite{abushawish2020pid}, thanks to its balance between convergence speed and solution quality \cite{kim2008robust, solihin2011tuning}. Genetic Algorithms (GA)~\cite{Goldberg1988GeneticAI} are among the earliest metaheuristic optimization techniques. Despite years of use, they remain widely applied in PID-controls problem \cite{jaen2013pid, meena2017genetic, suseno2021tuning} due to their versatility and robustness against local minima, making them a useful reference point for illustrating how metaheuristic methods have evolved over time. On the other hand, Grey Wolf Optimization (GWO) \cite{Mirjalili2014GreyWO}, a metaheuristic inspired by the social hierarchy and cooperative hunting strategy of grey wolves, has recently demonstrated competitive or superior performance in various applications \cite{Hashem2023GreyWO}. Its simplicity, efficient formulation, low computational cost, and adaptability have contributed to its growing use in engineering optimization and controller tuning problems \cite{madadi2014optimal, li2017enhanced}. These methods are gradient-free and well suited to non-linear, multi-modal cost landscapes. However, finding near-optimal solutions for high-dimensional problems with metaheuristics often requires many iterations of population updates or iterative searches, resulting in long computation times. In addition, they may need to be adapted for each specific problem \cite{wolpert2002no}.

Another family of optimization methods is represented by black-box optimization methods, such as Bayesian Optimization (BO) \cite{frazier2018bayesian}. Rather than evolving populations of candidate solutions as metaheuristics do, BO builds a probabilistic surrogate of the objective function and and guides the search through an acquisition function that balances exploration and exploitation. This strategy is particularly effective when evaluating candidate solutions is computationally expensive \cite{diessner2022investigating}, as in UAV simulations involving aerodynamics and acoustics. BO has shown strong performance in controller-tuning problems with a moderate number of decision variables, although its scalability to high-dimensional spaces is more limited \cite{santoni2024comparison, malu2021bayesian}. A key drawback is its strong dependence on the definition of the optimization domain: a poor choice may cause instability or slow convergence, thereby increasing the experimental cost. In practice, the domain is usually set using prior knowledge from simulations, experiments, or expert intuition. This reliance on manual tuning or trial-and-error approaches could limit the applicability of BO to automatic multi-loop PID controller tuning in multivariable processes where defining an appropriate design space is particularly difficult \cite{coutinho2023bayesian}.

More recently, Reinforcement Learning, and in particular Deep Reinforcement Learning, has emerged as a promising option for multiobjective tuning in continuous and high dimensional settings \cite{recht2019tour, li2020deep, zou2021reinforcement}. 
DRL has been applied to PID controller design, with studies reporting competitive results against classical methods and proposing various Markov Decision Process (MDP) formulations for gain tuning, including one-shot selection, online adaptive updating, and multi-phase schedules \cite{Carlucho2017IncrementalQL, Shi2018AdaptiveQL, Ding2023MultiPhaseDDPG, Chowdhury2023EMTD3}.

In adaptive gain tuning, the agent is a gain updater that adjusts PID parameters online; the state collects error signals and short history terms; the action is a new set of gains or gain increments; the reward measures tracking quality and control effort over a finite reference task \cite{cheng2007proposal,guan2021design}. Other works keep the agent online but focuses the state on error based features such as the error, its integral, and its rate of change; the action is an updated gain vector; the reward penalizes overshoot and oscillations during the run \cite{shuprajhaa2022reinforcement}. A different line embeds the PID relation in the decision itself: the agent outputs the control signal at each step, the state is the triplet of error, integral, and derivative, and the policy parameters play the role of effective gains learned across episodes \cite{bujgoi2024tuning}. In contrast, our study frames DRL tuning as a one step evolution: the agent proposes a complete PID vector, the environment returns the negative of a composite cost after a single simulation, and the episode ends immediately. This design is straightforward, essentially making the RL agent a black-box optimizer. This allows for a fair comparison with black-box optimizers while avoiding adaptive control during flight. However, it could also restrict the expressive power of the RL formulation. Since the resulting policies are static and lack temporal interaction, the agent may lose the ability to learn adaptive strategies that adjust gains in response to changing flight conditions.

This study presents a systematic comparison of three optimization approaches, metaheuristics, BO, and DRL, for tuning UAV PID controllers under realistic flight conditions. The proposed framework addresses multiple, often competing objectives, including flight stability, trajectory tracking accuracy, energy efficiency, and acoustic emissions. This contrasts with much of the existing literature, which typically evaluates optimization under narrow initial conditions or demonstrates controller performance through isolated responses or partial PID tuning \cite{kim2008robust, solihin2011tuning, jaen2013pid, george2022optimal}. To this end, we developed a comprehensive simulation environment that captures the coupled effects of aerodynamics, rigid-body dynamics, and environmental disturbances. The framework integrates several key components: a reduced-order aerodynamic model coupled with data-driven surrogates for computational efficiency, six degree-of-freedom rigid-body dynamics to simulate both translational and rotational motion, a stochastic turbulence model to replicate atmospheric disturbances, and an acoustic emission model to evaluate noise performance. The test scenario involves a quadcopter executing trajectory tracking maneuvers in turbulent conditions, with PID gains optimized through repeated closed-loop simulations.
Building on this methodological foundation, the paper proceeds to evaluate and compare the three optimization approaches within the established simulation framework. Section \ref{sec:method} provides comprehensive details on each framework component: UAV dynamics and aerodynamics, controller architecture, acoustic modeling, and the implementation of the three optimization algorithms. Section \ref{sec:results} presents a comparative analysis of the optimization strategies, examining their performance trade-offs and relative strengths across all objectives. Section \ref{sec:concl} synthesizes the key findings, acknowledges study limitations, and identifies promising avenues for future research.

\section{Methodology}\label{sec:method}

\begin{figure}
  \centering
  \makebox[\textwidth][c]{%
    \includegraphics[width=1.0\linewidth]{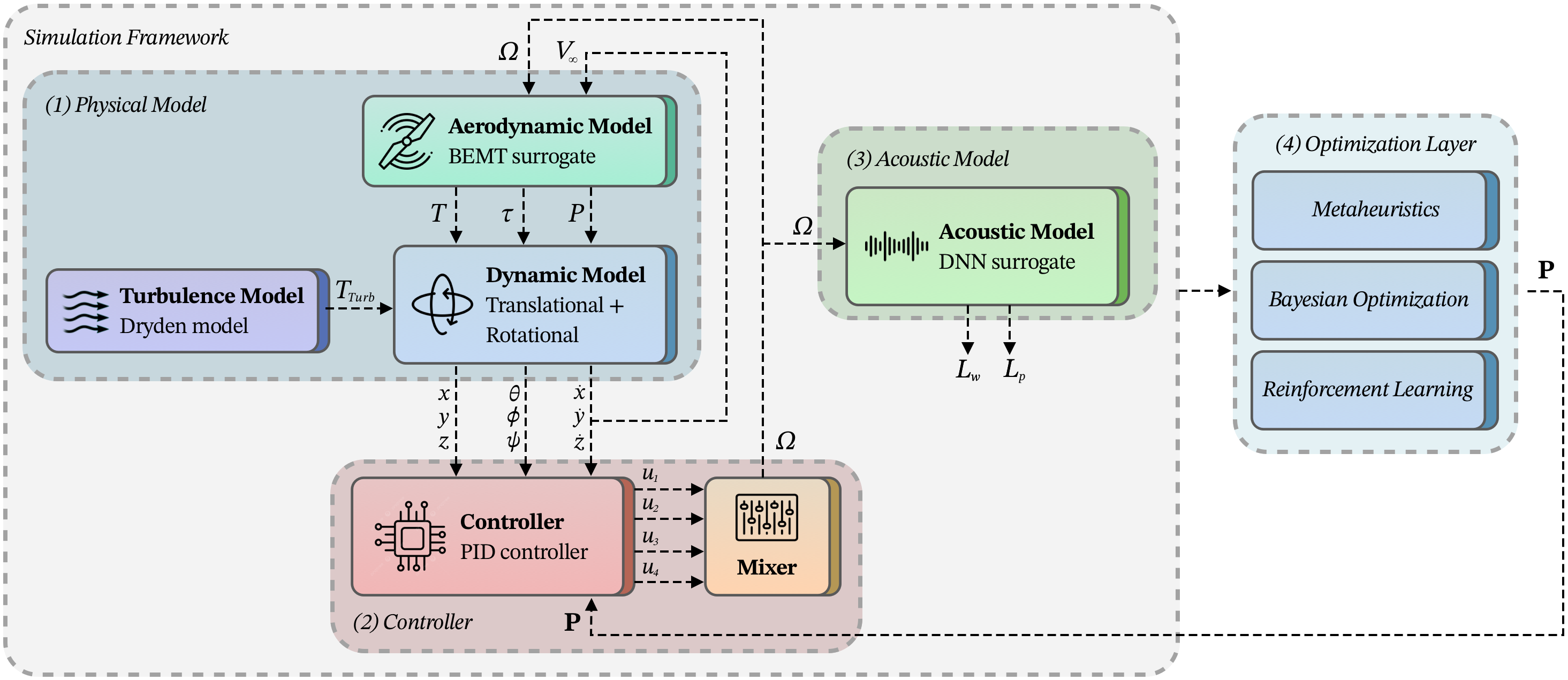} }
  \caption{Architecture component flowchart. The optimization layer output $\mathbf{P}$ represents a complete set of optimized PID.}
  \label{fig:flowchart}
\end{figure}

Our methodology aims to build a simulation environment that can cyclically and automatically evaluate the UAV performance by varying the PID configuration while generating useful synthetic data. The framework is specifically designed to incorporate acoustic emission metrics, alongside traditional flight dynamics, enabling the evaluation of control strategies not only in terms of stability, but also in terms of acoustic footprint and power efficiency. To achieve this, the following framework is structured into four principal modules, each conceived as an independent and interchangeable component: (1) a physical model, introduced in Section \ref{sec:phymodel}, that implements the dynamics and aerodynamics of the UAV in a simulated physical space, reproducing its rigid-body dynamics and rotor aerodynamics while incorporating turbulence effects; (2) a cascade PID-based controller responsible for translating navigation objectives into actuator commands, introduced in Section \ref{sec:controller}; (3) an acoustic model predicting noise emissions during the simulation, introduced in Section \ref{sec:acoustic}; and (4) an optimization layer, supporting metaheuristic algorithms (Section \ref{sec:meta}), BO (Section \ref{sec:bayesopt}) and DRL (Section \ref{sec:rl_onestep}) methods, tasked with tuning the PID gains to achieve an optimal trade-off between flight performance and noise mitigation. A flowchart of the entire architecture is reported in Figure \ref{fig:flowchart}.

The simulation framework is designed to establish a balance between simulation fidelity and computational efficiency for in-loop optimization. Higher-fidelity simulations generate more reliable evaluations without requiring costly real-world experiments for every candidate control configuration, however, they can quickly become computationally expensive, especially when aerodynamic forces are determined by complex simulation techniques, such as computational fluid dynamics (CFD). At the same time, a single PID-dependent simulation must be efficient enough to be effectively integrated into an optimization cycle.  This trade-off is particularly important for noise-aware control as well, since the acoustic footprint is influenced not only by steady-state rotor speeds but also by transient events, such as turbulence-induced load fluctuations. For this reason, the aerodynamic model needs to combine speed with accuracy, which is why data-driven surrogate models are often employed in real-time or accelerated simulation scenarios \cite{yondo2018review, sun2019review, immordino2025predicting}. To support reproducibility and adaptability to different fidelity requirements, the simulation environment has been implemented in Python as an open-source library. This allows for the integration of new aerodynamic models, acoustic models, controllers, and optimization algorithms that can easily be re-implemented or replaced with different reduced order models (ROM) or surrogates.


\subsection{Physical Model}\label{sec:phymodel}

\paragraph{\textbf{Blade Element Momentum Theory}}

While CFD can provide precise force estimation using iterative Navier-Stokes equation-based solvers, ROMs and low-order methods can offer a good balance of fidelity and efficiency. The aerodynamic model in our framework is based on the Blade Element Momentum Theory (BEMT) \cite{davoudi2019hybrid}, which combines blade element theory with momentum theory to compute rotor thrust and torque. These models are widely employed in aeroacoustic research~\cite{pacini2021towards} to construct flight-mechanics representations of drones, efficiently accounting for spanwise variations in angle of attack, chord length, and blade twist while maintaining low computational cost compared to traditional CFD. In our simulator, BEMT is used to generate rotor performance data (thrust $T_i$, torque $\tau_i$, power $P_i$ and corresponding coefficients) for different rotor speeds $\omega_i$ (see \ref{appx:bemt}) and freestream velocity. 

Despite its efficiency, BEMT remains a low-order method with important limitations. It does not capture three-dimensional flow effects such as tip vortices, dynamic stall, and strong rotor–wake interactions, which can strongly influence performance and noise predictions in realistic flight conditions \cite{boatto2023assessment}. Its use therefore represents a pragmatic compromise: it enables fast data generation within a reasonable time frame, but the results must be interpreted with an awareness of these simplifications.

\paragraph{\textbf{BEMT Surrogate}}
While BEMT provides good low-order predictions of rotor performance, repeated evaluation in an optimization loop can still be computationally expensive. To address this, a surrogate model based on a fully connected Deep Neural Network (DNN) has been trained offline using data generated from the implemented BEMT solver.
The inputs to the network are the rotor angular velocity $\Omega_i$ (RPM) and freestream velocity $V_{\infty}$ (m/s), and the outputs are six aerodynamic quantities: thrust $T_i$, torque $\tau_i$, power $P_i$, and their respective non-dimensional coefficients $C_T$, $C_Q$, and $C_P$.
The dataset was generated by sweeping rotor speed between 0 and 4,000~RPM and velocity from 1 to 20 m/s, yielding 8,000 samples divided into training and validation sets in an 80/20 split. All targets were normalized prior to training to improve convergence, and early stopping was employed to avoid overfitting.

To ensure robust evaluation, k-fold cross-validation with $k=5$ was performed, and the network architecture along with its hyperparameters was optimized using \texttt{optuna} within the PyTorch framework. After training, the DNN replaces the BEMT in the simulations, achieving a substantial reduction in computation time while maintaining adequate accuracy. Performance was evaluated using Symmetric Mean Absolute Percentage Error (SMAPE), Root Mean Square Error (RMSE), and Mean Absolute Error (MAE) metrics. The model exhibits a speedup of approximately $1,900\times$ with a SMAPE validation error of around $5.75\%$. The surrogate performance is outlined in Table ~\ref{tab:bemt_dnn_results}.

\begin{table}[H]
\centering
\small
\caption{DNN surrogate architecture, training configuration, and performance comparison with the BEMT solver.}
\label{tab:bemt_dnn_results}
\renewcommand{\arraystretch}{1.2}
\begin{tabular}{p{3.5cm} p{4.5cm} p{4cm}}
\hline
\textbf{ } & \textbf{Parameter} & \textbf{Value} \\ \hline
\multirow{4}{*}{\textbf{Architecture}} 
& Input features & 2 ($\Omega_i$, $V_{\infty}$) \\
& Hidden layers & 3 (16-32-16 units) \\
& Output features & 6 ($T_i, \tau_i, P_i, C_T, C_Q, C_P$) \\
& Activation function & ReLU \\ \hline
\multirow{5}{*}{\textbf{Training setup}}
& Loss function & MSE \\
& Optimizer & Adam \\
& Learning rate & $1\times 10^{-4}$ \\
& Training-set size & 6{,}400 \\
& Validation-set size & 1{,}600\\ \hline
\multirow{6}{*}{\textbf{Performance}}
& Training time & $\approx40$ s per fold \\
& DNN Prediction time & $1.46\times 10^{-4}$ s \\
& BEMT Prediction time & $2.7\times 10^{-1}$ s \\
& MAE & 3.18 \\
& RMSE & 8.99 \\
& SMAPE & 5.75\% \\ \hline
\multicolumn{2}{l}{\textbf{Speedup (BEMT/DNN)}} & $\approx 1{,}900\times$ \\
\hline
\end{tabular}
\end{table}

An important feature of the surrogate implementation is model interchangeability.
The simulator queries rotor forces and moments through a unified interface, meaning that the DNN can be retrained with datasets obtained from higher-fidelity simulations or experimental rotor measurements. Alternatively, it can be substituted with more advanced neural architectures without altering the rest of the simulation framework. This design choice directly supports the integration of Navier-Stokes surrogates based on promising new ML approaches, which have recently demonstrated strong potential and applicability \cite{massegur2024graph, immordino2025predicting, immordino2025spatio, vaiuso2025multi}.
Modularity enables the aerodynamic model to evolve in fidelity without compromising the efficiency necessary for large-scale optimization, thus achieving an adaptable balance between computational cost and physical accuracy.

\begin{figure}
    \centering
    \includegraphics[width=1\linewidth]{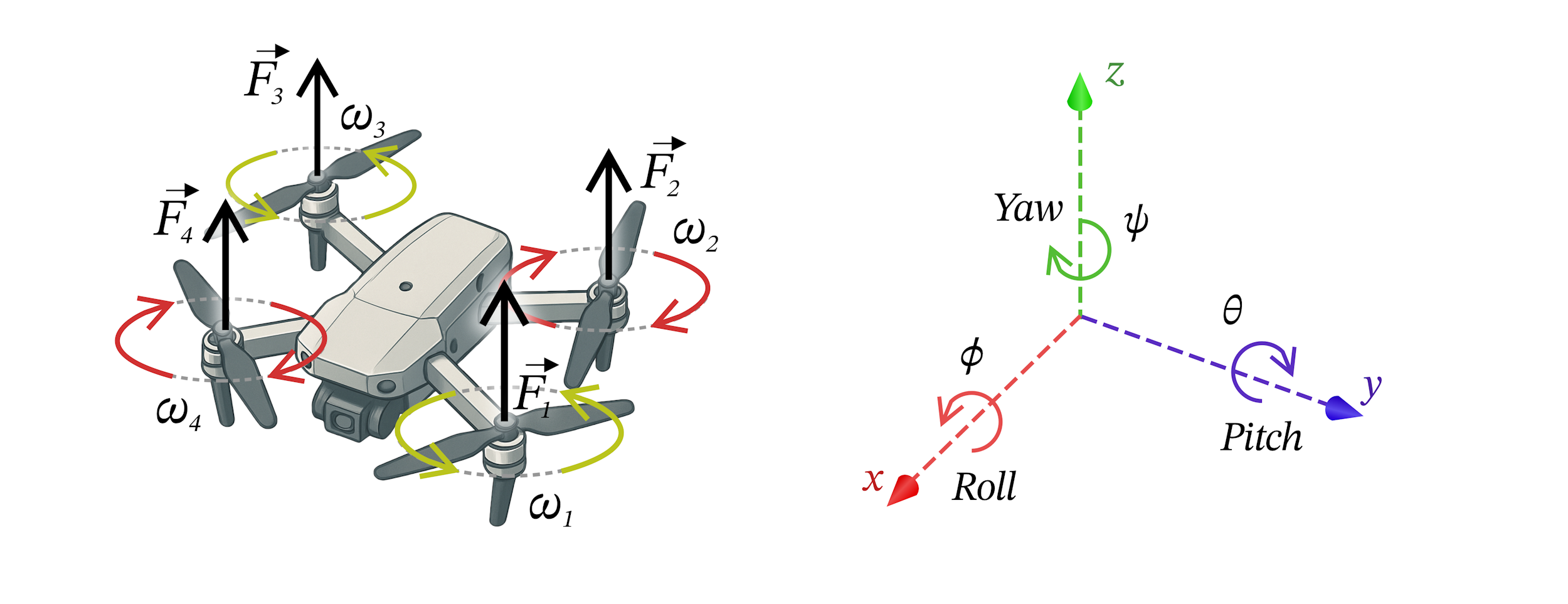}
    \caption{Physical quadricopter drone model representation}
    \label{fig:drone}
\end{figure}

\paragraph{\textbf{Translational dynamics}}
Translational and rotational UAV dynamics are simulated using a rigid-body six degree-of-freedom model, represented in Figure \ref{fig:drone}, with roll ($\phi$), pitch($\theta$), and yaw ($\psi$) correspond to rotations about the body x, y, and z axes. We follow the formulation in~\cite{abdelhay2019modeling}, adapted to use thrust and torque values computed from the BEMT surrogate implementation.  
The translational motion law is:
\begin{equation}
\label{eq:transl}
m\,\dot{\mathbf{v}} \;=\; \mathbf{F}_T + \mathbf{F}_D - m\,\mathbf{g},
\end{equation}
where $m$ is the UAV mass, $\mathbf{v} = [\dot{x}, \dot{y}, \dot{z}]^\top$ is the velocity in the world frame, and $\mathbf{g}$ is the gravity acceleration vector.  

The total thrust force in the world frame is:
\[
\mathbf{F}_T = \mathbf{R}(\phi, \theta, \psi)\,[0,0,T_{\Sigma}]^\top
\]
where $T_{\Sigma} = \sum_{i=1}^4 T_i$ is the sum of the four rotor thrusts and $\mathbf{R}(\phi, \theta, \psi)$ is the body-to-world rotation matrix computed with a $z$–$y$–$x$ Euler sequence, using roll $\phi$, pitch $\theta$, and yaw $\psi$.  

The aerodynamic drag is modeled as:
\[
\mathbf{F}_D = - \mathrm{diag}(C_{d,x}, C_{d,y}, C_{d,z}) \, \mathbf{v},
\]
where $C_{d,x}$, $C_{d,y}$, and $C_{d,z}$ are linear drag coefficients along each axis.  

Expanding Equation~\eqref{eq:transl} for each component yields:
\begin{align}
\ddot x &= \frac{T_{\Sigma}}{m}\big(\cos\psi\,\sin\theta\,\cos\phi + \sin\psi\,\sin\phi\big)
          - \frac{C_{d,x}}{m}\,\dot x, \\
\ddot y &= \frac{T_{\Sigma}}{m}\big(\sin\psi\,\sin\theta\,\cos\phi - \cos\psi\,\sin\phi\big)
          - \frac{C_{d,y}}{m}\,\dot y, \\
\ddot z &= \frac{T_{\Sigma}}{m}\big(\cos\theta\,\cos\phi\big)
          - \frac{C_{d,z}}{m}\,\dot z - g.
\end{align}

In-time state propagation uses a fixed-step, fourth-order Runge--Kutta scheme \cite{Jameson1981NumericalSO} (see \ref{appx:rks}). This integrator offers a good trade-off between accuracy and cost for closed-loop optimization runs and is standard in flight-dynamics simulation \cite{ando2025ExploringER}.

\paragraph{\textbf{Rotational dynamics}}
The UAV rotational motion is governed by the Euler rigid-body equations with gyroscopic coupling.  
Let $\mathbf{I} = \mathrm{diag}(I_x, I_y, I_z)$ be the inertia matrix, $p = \dot{\phi}$, $q = \dot{\theta}$, $r = \dot{\psi}$ the angular velocity components in the body frame, and $J_r$ the combined spinning inertia of the rotor and shaft.  

Control moments are decomposed along the body axes as:
\begin{align}
u_1 &= T_{\Sigma} \quad \text{(total thrust)}, \\
u_2 &= l\,(T_4 - T_2) \quad \text{(roll moment)}, \\
u_3 &= l\,(T_3 - T_1) \quad \text{(pitch moment)}, \\
u_4 &= \tau_1 - \tau_2 + \tau_3 - \tau_4 \quad \text{(yaw moment)},
\end{align}
where $l$ is the arm length from the center to each rotor, and $\tau_i$ are the rotor reaction torques computed from the BEMT surrogate.  

Let $\Omega_{\mathrm{diff}} = (\omega_1 - \omega_2 + \omega_3 - \omega_4)$ be the signed sum of rotor angular speeds (counter-rotating pairs). The rotational dynamics then read:
\begin{align}
I_x \dot p &= u_2 - C_{a,x}\,\mathrm{sgn}(p)\,p^2 - J_r \,\Omega_{\mathrm{diff}}\, q - (I_z - I_y)\,q\,r, \\
I_y \dot q &= u_3 - C_{a,y}\,\mathrm{sgn}(q)\,q^2 + J_r \,\Omega_{\mathrm{diff}}\, p - (I_x - I_z)\,p\,r, \\
I_z \dot r &= u_4 - C_{a,z}\,\mathrm{sgn}(r)\,r^2 - (I_y - I_x)\,p\,q,
\end{align}
where $C_{a,x}$, $C_{a,y}$, $C_{a,z}$ are quadratic rotational damping coefficients.

\paragraph{\textbf{Dryden turbulence thrust augmentation}}
Atmospheric turbulence is modeled using the Dryden spectral model~\cite{dryden1943review}, a stochastic process representation frequently adopted in flight simulation and control system evaluation (see ~\ref{appx:dryd}). Stochastic wind components $(u_{\mathrm{turb}}, v_{\mathrm{turb}}, w_{\mathrm{turb}})$ represent, respectively, the turbulent velocities along the $x$, $y$, and $z$ axes of the inertial frame. These components are projected onto the rotor disk to obtain the local relative velocity:
\[
U_{\mathrm{disk}}(\theta,r) = \omega \, r + u_{\mathrm{turb}} \cos\theta + v_{\mathrm{turb}} \sin\theta
\]
where $\omega$ is the rotor angular velocity (rad/s), $r$ is the radial position along the blade span, and $\theta$ is the azimuthal angle around the disk.  

The instantaneous load density over the disk is expressed as:
\[
f(\theta,r) = \frac{2\pi\, w_{\mathrm{turb}}}{\omega\, r} \, U_{\mathrm{disk}}^2(\theta,r),
\]

Integrating over the rotor annulus, with $R_1$ and $R_2$ denoting the inner and outer radii (root and tip) of the blade Section considered, yields:
\begin{equation}
\label{eq:dryden_int}
\int_0^{2\pi}\!\!\int_{R_1}^{R_2} f(\theta,r)\,dr\,d\theta
= 2\pi^2 w_{\mathrm{turb}} \omega \,(R_2^2 - R_1^2)
  + \frac{2\pi^2 w_{\mathrm{turb}}}{\omega}\,(u_{\mathrm{turb}}^2+v_{\mathrm{turb}}^2)\,\ln\!\frac{R_2}{R_1}.
\end{equation}

The turbulence-induced thrust increment is then:
\begin{equation}
T_{\mathrm{turb}} = \tfrac{1}{2}\rho
\left[ \int_0^{2\pi}\!\!\int_{R_1}^{R_2} f(\theta,r)\,dr\,d\theta \right]
\end{equation}
\begin{equation}
\label{eq:thrust_tot}
    T_{\mathrm{tot}} = T_{\Sigma} + T_{\mathrm{turb}}
\end{equation}

Equation~\eqref{eq:dryden_int} is evaluated for each rotor individually and its contribution is added to the BEMT-computed thrust following Equation~\eqref{eq:thrust_tot} .

\subsection{Controller}\label{sec:controller}

The UAV is stabilised using a cascade PID architecture inspired by open-source flight stacks PX4 \cite{meier2015px4}. The structure consists of three main control loops:
\begin{itemize}
    \item \textbf{Outer loop (position \& altitude control):} Computes desired horizontal and vertical velocities from position and altitude errors.  
    \item \textbf{Middle loop (attitude control):} Converts desired velocities into roll and pitch angle setpoints, and generates angular rate setpoints for roll, pitch, and yaw.  
    \item \textbf{Inner loop (rate control):} Converts the angular rate setpoints to produces torque and thrust commands, which are finally distributed to the individual rotors via the mixer.  
\end{itemize}

\begin{figure}
    \centering
    \includegraphics[width=1\linewidth]{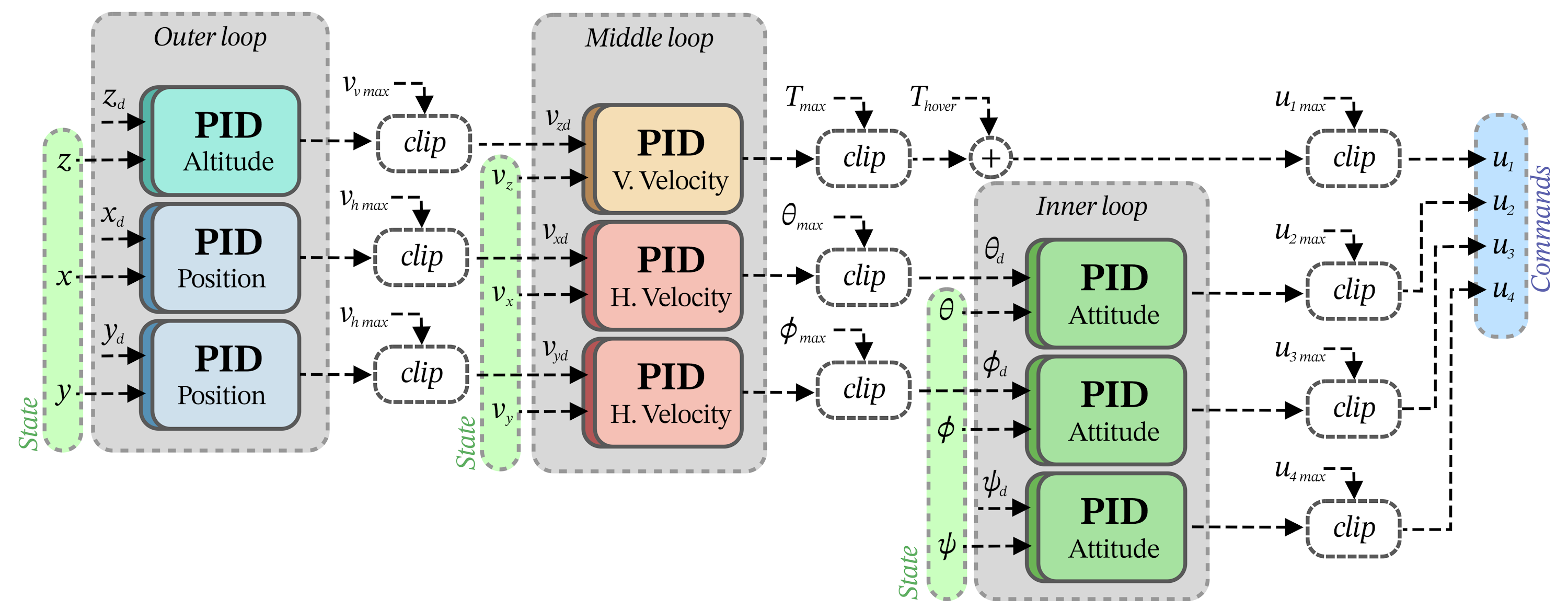}
    \caption{Representation of the quadcopter controller layout. Dashed green boxes indicate if a parameter is part of the drone current state, while $z_d$, $x_d$, $y_d$, $v_{zd}$, $v_{xd}$, $v_{yd}$, $\theta_d$, $\phi_d$ and $\psi_d$ are the desired values calculated based on target position.}
    \label{fig:pid_layout}
\end{figure}

\paragraph{\textbf{Layout}}
In total, the system implements five independent PID controllers: position in the $x$ and $y$ axes (sharing the same gain set), altitude control, roll and pitch and yaw attitude control (sharing the same gain set), horizontal speed control, and vertical speed control. Each PID block has proportional, integral, and derivative gains $\{K_p, K_i, K_d\}$, resulting in 15 possible tunable parameters. The whole layout is represented in Figure \ref{fig:pid_layout}. The outer loop regulates position and altitude, generating desired velocities. The middle loop maps velocities to desired thrust and attitudes. The inner loop tracks attitudes, outputting per-rotor commands $u_1$–$u_4$. Some PID controllers share the same gains to reduce the dimensionality of the optimization problem and to reflect inherent system symmetries. For instance, the UAV dynamics in the $x$ and $y$ directions are nearly identical under nominal conditions, which justifies adopting a common gain set for both position axes. Similarly, roll and pitch and yaw exhibit comparable dynamic behavior due to the geometric symmetry of the quadcopter frame, allowing them to be stabilized effectively with the same set of attitude gains. This parameter grouping not only simplifies tuning but also improves optimization efficiency by reducing the search space without sacrificing control performance.

Several mechanisms are included to improve stability and robustness. The controllers use anti-windup \cite{Galeani2009ATO} by clamping the integral term to a limit proportional to the control authority, avoiding integral runaway during large or sustained errors. Command saturation is applied individually to thrust, roll, pitch, and yaw outputs, ensuring they remain within physical limits. Desired horizontal and vertical speeds, as well as roll and pitch angles, are bounded to prevent aggressive maneuvers that could destabilize the UAV. A feed-forward hover compensation term adjusts thrust commands according to the instantaneous roll and pitch, maintaining equilibrium during maneuvers \cite{meier2015px4}.

\paragraph{\textbf{Mixer and thrust/torque mapping}}  
The mixer converts the desired collective thrust and body-axis moments into individual rotor speeds, taking into account quadrotor geometry, rotor spin directions, and the thrust--torque characteristics of each propeller--motor unit.  

The control inputs to the mixer are:
\[
u_1 \; \text{(collective thrust)}, \quad
u_2 \; \text{(roll moment)}, \quad
u_3 \; \text{(pitch moment)}, \quad
u_4 \; \text{(yaw moment)} .
\]  

For a quadrotor in the ``$+$'' configuration (where one rotor is aligned with the forward axis of the UAV), the mapping from rotor thrusts $\mathbf{T} = [T_1, T_2, T_3, T_4]^\top$ to control inputs, as defined in \cite{abdelhay2019modeling}, is:
\[
\begin{bmatrix}
u_1 \\[2pt]
u_2 \\[2pt]
u_3 \\[2pt]
u_4
\end{bmatrix}
=
\underbrace{
\begin{bmatrix}
1 & 1 & 1 & 1 \\[2pt]
0 & -l & 0 & l \\[2pt]
-l & 0 & l & 0 \\[2pt]
\frac{d}{C_T} & -\frac{d}{C_T} & \frac{d}{C_T} & -\frac{d}{C_T}
\end{bmatrix}
}_{\mathbf{A}}
\begin{bmatrix}
T_1 \\[2pt]
T_2 \\[2pt]
T_3 \\[2pt]
T_4
\end{bmatrix},
\]  
where $C_T$ is the thrust coefficient and $d$ is the drag factor. The inverse mapping distributes the control inputs into individual rotor thrusts:  
\[
\mathbf{T} = \mathbf{A}^{-1} \,
\begin{bmatrix}
u_1 \\[2pt]
u_2 \\[2pt]
u_3 \\[2pt]
u_4
\end{bmatrix}.
\]  

Using the relation $T_i = C_T \, \omega_i^2$, this can be expressed directly in terms of rotor angular speed squared:  
\[
\begin{bmatrix}
\omega_1^2 \\[2pt]
\omega_2^2 \\[2pt]
\omega_3^2 \\[2pt]
\omega_4^2
\end{bmatrix}
=
\frac{1}{4 C_T}
\begin{bmatrix}
1 & 0 & -2/l & d/C_T \\[2pt]
1 & -2/l & 0 & -d/C_T \\[2pt]
1 & 0 & 2/l & d/C_T \\[2pt]
1 & 2/l & 0 & -d/C_T
\end{bmatrix}
\begin{bmatrix}
u_1 \\[2pt]
u_2 \\[2pt]
u_3 \\[2pt]
u_4
\end{bmatrix}.
\]  

The mixer ensures physical feasibility by clipping $\omega_i^2$ to the range $[0,\,\omega_{\max}^2]$. The rotor speeds are then obtained from square roots and converted to revolutions per minute (RPM). The aerodynamic model then maps $(\mathrm{RPM}_i)$ to thrust $T_i$ and torque $\tau_i$ for each rotor, coupling the control inputs to the rigid-body dynamics of the UAV.

\subsection{Acoustic Model}\label{sec:acoustic}

The acoustic emissions of the multi-rotor UAV are modeled as a function of the radiation angle $\zeta$ and the rotor rotational speed $\Omega$, since multicopter noise depends both on rotor RPM which control blade passing frequency (BPF), and on a characteristic directivity pattern with stronger radiation downward than sideways. In general, as the RPM increases, the overall sound power rises and the spectral content shifts toward higher frequencies. For a given operating condition, the sound power level spectrum $L_w(f, \zeta, \Omega)$ is expressed in discrete 1/3-octave bands $f_i$~\cite{international2014iec}. The use of 1/3-octave bands provides sufficient resolution to capture the relevant tonal components of rotor noise while avoiding the complexity and data requirements of narrowband analyses such as FFT \cite{ramos2022requirements}, though it is not adequate for detailed psychoacoustic evaluation \cite{wunderli2022method}. This representation offers a practical balance between spectral fidelity and computational tractability for UAV noise modeling.

To integrate an acoustic model into the simulation framework, different approaches were evaluated, with the goal of ensuring high accuracy while maintaining computational efficiency for use in optimization loops. Data-driven methods were selected as the most suitable compromise, as they can capture complex acoustic dependencies while remaining faster than high-fidelity physics-based models \cite{thurman2023comparison, poggi2022neural}. Among existing approaches, the empirical regression model proposed by Wunderli~\cite{wunderli2022method} was initially implemented. This method assumes that the directivity pattern of the UAV noise is nearly independent of rotor speed and flight maneuver, expressing $L_w$ as a polynomial function of the deviations in radiation angle $\Delta \zeta$ and rotor speed $\Delta \Omega$ from a reference condition (e.g., hover), plus a maneuver-dependent correction term $C_{\text{proc}}$.  

A second approach employed a fully connected DNN to directly regress the frequency-domain spectrum. Both models were trained on a real data measurement dataset \cite{righi_2024_10512044}, acquired according to ISO 5305 \cite{ISO5305_2024} standards during a campaign on a DJI Matrice 300 UAV~\cite{vaiuso2024drone}, yielding 210 samples containing log data on $\Omega$ (RPM), $\zeta$ (rad), pitch, roll, yaw, and $C_{\text{proc}}$, calculated following \cite{wunderli2022method}, and the third-octave-band $L_w$ spectrum used as supervised label.  

For the polynomial model, after normalizing the RPM and $C_{\text{proc}}$ values, the regression converged in 55 iterations with a runtime of about 3 minutes.

The surrogate archived a SMAPE of $9.89\%$, RMSE of $7.56$, and MAE of $5.59$ on a separated test set of 32 samples. The DNN, consisting of three fully connected layers with LeakyReLU activation, was trained using the Adam optimizer with a learning rate of $10^{-4}$. Early stopping and hyperparameter optimization (\texttt{optuna}) were used to prevent overfitting, with performance evaluated via k-fold cross-validation with $k=5$. The dataset was split into 147 training, 31 validation, and 32 test samples. The optimized DNN outperformed the empirical model, achieving a test SMAPE of $3.29\%$, RMSE of $2.5$, and MAE of $1.9$, with an average training time of 30 seconds per fold and $\approx27$ times faster prediction time.  

For simulation, the trained DNN is combined with a lookup table of angle-dependent corrections to compute the instantaneous sound power level for each rotor at its current orientation. The contribution of a single rotor is isolated by subtracting $6~\mathrm{dB}$ from the reference hovering spectrum. 

\begin{table}[H]
\centering
\small
\caption{Comparison of the Wunderli regression model~\cite{wunderli2022method} and the optimized DNN for UAV acoustic prediction.}
\label{tab:acoustic_model_results}
\renewcommand{\arraystretch}{1.2}
\begin{tabular}{p{3cm} p{3.7cm} p{3cm} p{3cm}}
\hline
\textbf{} & \textbf{Parameter} & \textbf{Wunderli Model \cite{wunderli2022method}} & \textbf{DNN Model} \\ \hline

\multirow{4}{*}{\textbf{Architecture}}
& Input features & 6 ($\Omega$, $\zeta$, $C_{\text{proc}})$ & 5 ($\Omega$, $\zeta$) \\
& Hidden layers & $2^\text{nd}$ order poly. & 3 (13-40-63 units) \\
& Output features & 63 ($L_w$) & 63 ($L_w$) \\
& Activation function & -- & LeakyReLU \\ \hline

\multirow{6}{*}{\textbf{Training setup}}
& Loss function & MSE & MSE \\
& Optimizer & L-BFGS-B & Adam \\
& Learning rate & -- & $1\times 10^{-4}$ \\
& Training-set size & 178 & 147 \\
& Validation-set size & -- & 31 \\
& Test-set size & 32 & 32 \\ \hline

\multirow{5}{*}{\textbf{Performance}}
& Training time & 206 s & $\approx 30$ s per fold \\
& Prediction time & $3.5\times 10^{-3}$ & $1.3\times 10^{-4}$ \\
& MAE & 5.59 & 1.90 \\
& RMSE & 7.56 & 2.50 \\
& SMAPE & 9.89\% & 3.29\% \\
\hline
\multicolumn{2}{l}{\textbf{Speedup (Wunderli/DNN)}} & \multicolumn{2}{c}{$\approx 27\times$} \\
\hline

\end{tabular}
\end{table}

\subsubsection*{Propagation Model}
\begin{figure}
    \centering
    \begin{subfigure}[b]{0.48\linewidth}
        \centering
        \includegraphics[width=\linewidth]{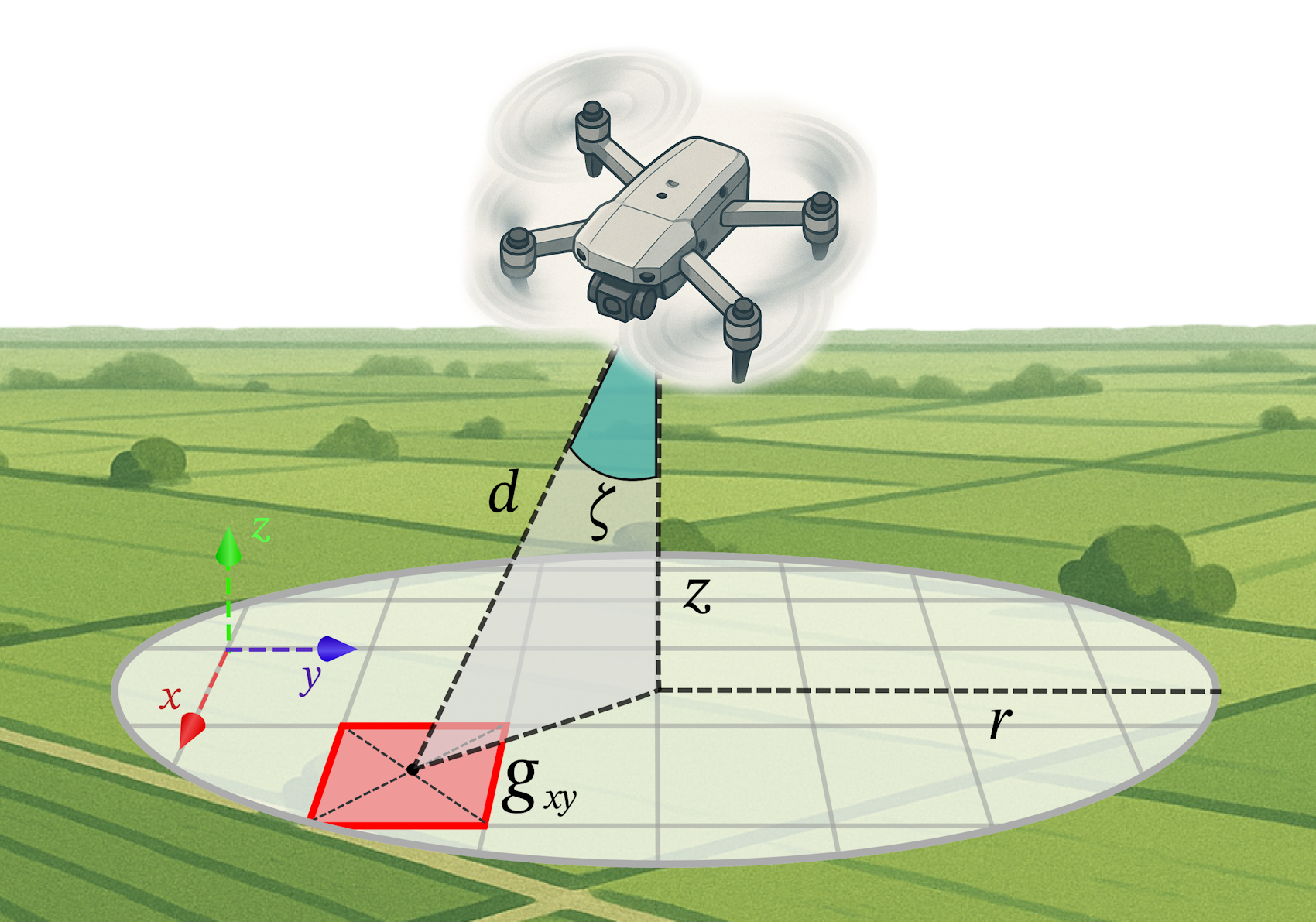}
        \caption{}
        \label{fig:noise_schema}
    \end{subfigure}
    \hfill
    \begin{subfigure}[b]{0.48\linewidth}
        \centering
        \includegraphics[width=\linewidth]{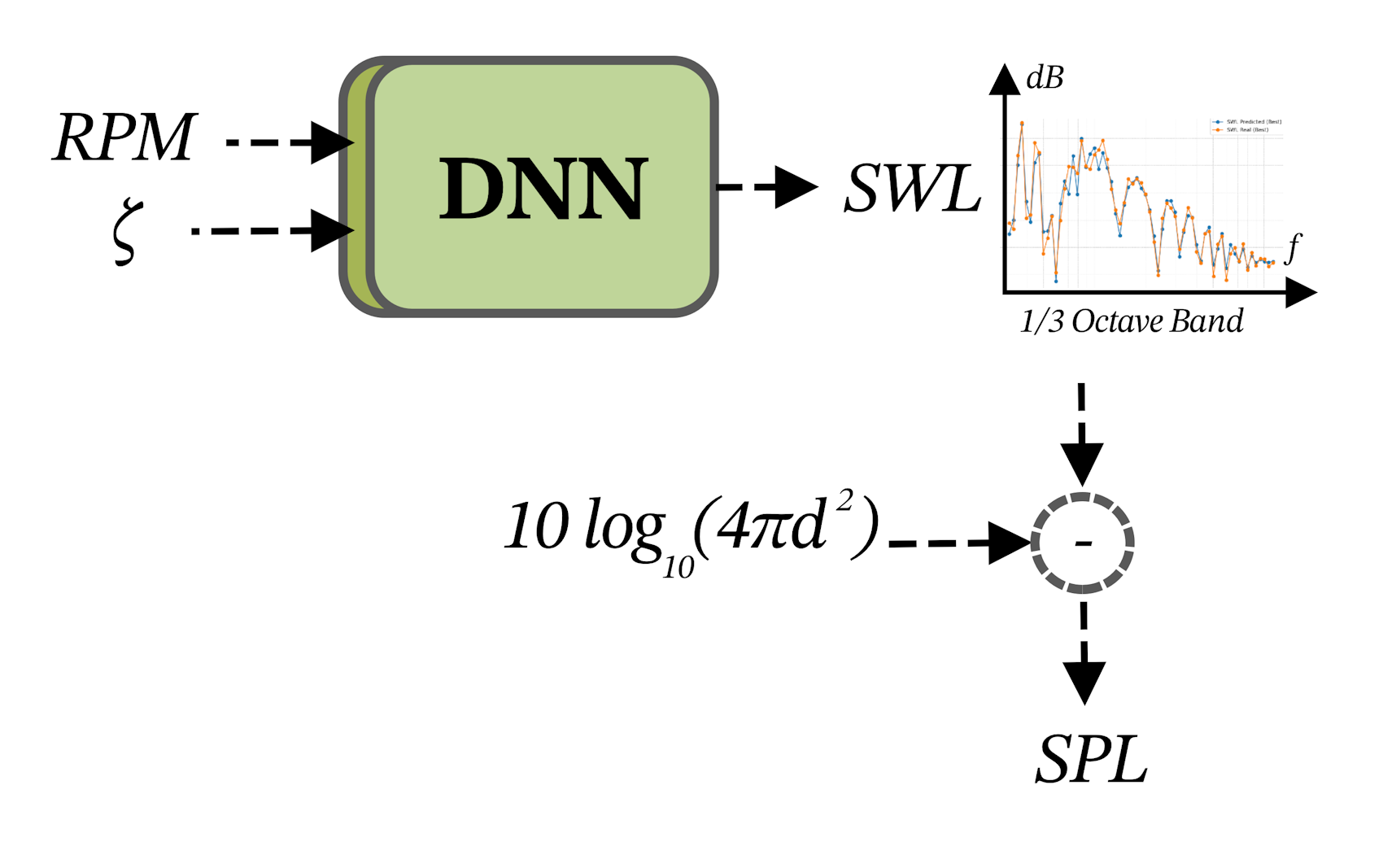}
        \caption{}
        \label{fig:noise_model}
    \end{subfigure}
    \caption{Figure (a) Discretized simulation space domain in cells. Figure (b) Schematic of the noise model that generates the final SPL value.}
    \label{fig:noise_combined}
\end{figure}

In order to calculate the step-by-step Sound Pressure Level (SPL) signal in the ground, $L_w$ is propagated to ground receivers using a free-field spherical spreading model with optional ISO~9613-1 \cite{ISO9613_1_1993} atmospheric absorption. 
The simulation space domain is discretized into a spatial $N\times N$ grid, with each grid cell $g$ represented by its horizontal centroid at height $z=0$. At each simulation time step $t_j$, the UAV position and orientation determine the instantaneous distance $d_g(t_j)$ and radiation angle $\zeta_g(t_j)$ for each cell, as shown in Figure \ref{fig:noise_schema}. The sound pressure level signal is aggregated step by step within each cell at a radius limit of $r$, using the radiation angle $\zeta$ and the distance from the cell centroid to the noise source of $d$. Using the propagation model above, the 1/3-octave-band SPL spectrum $L_p(f, t_j, g)$ is computed for all grid points, following the approach illustrated in Figure \ref{fig:noise_model}.

Propagation from the UAV to each receiver point is computed on a per-band basis. The free-field spherical spreading attenuation is:
\begin{equation}
A_{\mathrm{sp}}(d) = 10 \log_{10} \left( 4 \pi d^2 \right)
\end{equation}

Atmospheric absorption per frequency band is computed according to ISO~9613-1 \cite{ISO9613_1_1993}: $Atm_{abs}(f,f_{\mathrm{rO}},f_{\mathrm{rN}},p_r,T,T_0)$ as a function of $f_{\mathrm{rO}}$ and $f_{\mathrm{rN}}$ that are the oxygen and nitrogen relaxation frequencies respectively, $p_r$ is the relative pressure, $T$ is the absolute temperature (K), and $T_0 = 293.15\,\mathrm{K}$. An optional directivity index $DI(f, \theta)$ can be added if not already embedded in the reference emission spectra (see the complete formulation in \ref{appx:atm_direct}). 

The received SPL per band at a receiver point is then:
\begin{equation}
L_p(f, t) = L_w^{\mathrm{total}}(f, t) - A_{\mathrm{sp}}(d(t)) - Atm_{abs}(f) \, d(t) + DI(f, \theta(t))
\end{equation}

To reduce computational complexity, spectral updates are skipped for cells whose predicted broadband SPL is outside a pre-defined radius $r$ from the projected UAV position in the ground (for reproducibility, simulation parameters are reported in \ref{appx:param}).

\subsection{Optimization Framework}\label{sec:opt_frw}

The optimization framework is designed to evaluate candidate PID gain configurations by simulating the UAV response in predefined mission scenarios and computing a composite cost function that incorporates multiple performance metrics.

To determine the optimal PID controller gains for UAV control, we employ three main families of gradient-free optimization approaches:

\begin{itemize}
    \item \textbf{Metaheuristic search algorithms}, which explore the parameter space by evolving a population of candidate solutions according to heuristic rules inspired by natural processes, which iteratively update a population of candidate solutions based on exploration–exploitation heuristics.  
    \item \textbf{Bayesian Optimization (BO)}, which builds a probabilistic surrogate of the objective function and selects new candidates by maximizing an acquisition function that balances exploration and exploitation. 
    \item \textbf{Reinforcement Learning (RL) methods}, where an agent learns an optimal control policy by interacting with the simulation environment and receiving scalar feedback (rewards) that encode performance objectives.  
\end{itemize}

All strategies share a common evaluation framework: each candidate set of gains is tested in the UAV simulation, the resulting trajectory, energy consumption, and acoustic profile are recorded, and a composite cost function is computed, ensuring a fair comparison between algorithms under identical conditions.

The evaluation of each candidate PID configuration is performed by simulating the UAV in a fixed representative mission scenario using the simulation framework, and computing the cost function. To avoid overfitting the gains to a single route or a short path, we construct a single composite mission $\tilde s=(s_1,\dots,s_J)$ that concatenates heterogeneous segments (short hops, long transits, S-turns, climb/descents, hover–translate–hover) with varied speeds and altitudes. Additionally, the introduction of the Dryden turbulence, can generate noise, making the overall optimization more robust and generalizable. 

This approach diverges from existing literature, where optimization performance is typically evaluated under limited initial conditions or demonstrates controller effectiveness through single responses or individual PID component tuning \cite{kim2008robust, solihin2011tuning, jaen2013pid, george2022optimal}. Given the inherent nonlinear coupling characteristics of the quadrotor system, comprehensive evaluation requires challenging the controller with complex, multi-dimensional trajectories that exercise the full range of system dynamics. The selected trajectory design employs heuristic principles that systematically explore the vehicle flight envelope, ensuring that the optimization algorithms are tested across representative operational conditions covering most of the flight envelope.

The cost function is designed to penalize long mission times, large final position errors, excessive oscillations in attitude and thrust, high energy consumption, waypoint overshoot, incomplete missions, and elevated acoustic levels. Each term is weighted to reflect its relative importance in the specific application, with gains determined through an heuristic tuning process. From a scientific perspective, the exact choice of these weights is not critical, as it is application-specific and primarily serves to balance competing objectives. The cost function weights were calibrated by first obtaining a manually tuned baseline controller using the Ziegler–Nichols method \cite{meshram2012tuning}. To prevent unnecessary computation, simulations were terminated early when PID gains produced clearly unfeasible behavior. Specifically, runs were aborted if the UAV rapidly diverged from the initial waypoint, failed to initiate movement, or required excessive time to reach the first target. In such cases, a high penalty cost was assigned. Individual cost terms were then balanced to comparable magnitudes under this reference configuration, ensuring that no single metric disproportionately dominated the optimization. The resulting cost function promotes fast, accurate, stable, energy-efficient, and acoustically unobtrusive flight behavior.

The total cost is represented by the following equation, where each term is explained in Table \ref{tab:cost_table}.
\begin{equation} \label{eq:cost}
    J(\boldsymbol{P}) = w_t\,C_t + w_d\,C_d + w_o\,C_o + w_c\,C_c + w_{os}\,C_{os} + w_p\,C_p + w_n\,C_n + C_{nm}
\end{equation}

where $\boldsymbol{P}\in\mathbb{R}^{d}$ stacks the controller gains to be optimized.

\begin{table}[H]
\centering
\scriptsize
\caption{Cost function components}
\label{tab:cost_table}
\renewcommand{\arraystretch}{1.2}
\begin{tabular}{m{3.5cm} m{4cm} p{6.3cm}}
\hline
\textbf{Penalty term} & \textbf{Formula} & \textbf{Description} \\ \hline
 Mission time cost ($C_t$) & $T_f$ & Total simulation time required to reach the final waypoint. Lower values indicate faster mission completion. \\
Final position error ($C_d$)  & $\|\mathbf{p}_f - \mathbf{p}^\ast\|^{\gamma_d}$ & Euclidean distance between the UAV’s final position $\mathbf{p}_f$ and the target $\mathbf{p}^\ast$, shaped by exponent $\gamma_d\in(0,1]$ to control sensitivity to large errors in missions where the final target is not reached. \\
 Attitude oscillation ($C_o$) & $\sum_k(|\Delta\phi_k| + |\Delta\theta_k|)$ & Sum of absolute changes in roll ($\phi$) and pitch ($\theta$) over all time steps $k$, representing unnecessary or unstable attitude variations. \\
 Thrust oscillation ($C_{to}$)& $\sum_k |\Delta T_k|$ & Sum of absolute variations in total thrust command $T_k$ over the mission, penalizing rapid fluctuations in rotor speed demand. \\
 Completion cost ($C_c$)& $\begin{cases}0,&\text{mission completed}\\ P_c,&\text{otherwise}\end{cases}$ & Large increasing penalty $P_c$ applied if the UAV fails to visit all waypoints within the allowed time. \\
 Overshoot ($C_{os}$) & $\sum_{j=1}^{N_w} \max(0, \|\mathbf{p}_{\max,j} - \mathbf{p}^\ast_j\|)$ & Penalizes overshooting each waypoint $j$ by measuring the maximum excess distance $\mathbf{p}_{\max,j}$ beyond the target $\mathbf{p}^\ast_j$. \\
 Power consumption ($C_p$)& $\sum_k P_k$ & Total electrical energy expenditure computed as the sum of instantaneous power draws $P_k$ at each time step. \\
 Noise cost ($C_n$) & $\|\mathbf{s}\|_p^p + \max(\mathbf{s})$ & Combination of the $p$-norm of the broadband SWL vector $s$, obtained by power-summing the per-band $L_w$ values across frequency, and its maximum value $\max(\mathbf{s})$. \\
 No-movement cost ($C_{nm}$) & $\begin{cases}P_{nm},&\text{if disp.}<\epsilon\\ 0,&\text{otherwise}\end{cases}$ & Large penalty $P_{nm}$ if the UAV displacement is below the threshold $\epsilon$, preventing and explicitly skipping PID solutions that underestimate the minimum power required to initiate and sustain UAV motion. \\
\hline
\end{tabular}
\end{table}

The $w_{\cdot}$ coefficients are application-specific weights obtained heuristically to balance the influence of each term in the aggregated objective.

\subsection{Metaheuristic Optimization}\label{sec:meta}

We cast PID tuning as the minimization of the black–box objective \(J(\boldsymbol{P})\) in \eqref{eq:cost}. A metaheuristic maintains a population \(\{\boldsymbol{P}_i\}_{i=1}^{N}\) of candidate gain vectors, evaluates \(J(\boldsymbol{P}_i)\) by closed–loop simulation, and updates the population using exploration–exploitation rules that do not require gradients. The formulations below follow established practice for controller tuning \cite{abushawish2020pid}.

\subsubsection*{Genetic Algorithm}

In a generational Genetic Algorithm (GA) \cite{Goldberg1988GeneticAI}, each candidate \(\boldsymbol{P}_i\) is assigned a fitness \(F_i\) monotonically related to performance, e.g.\ \(F_i=-J(\boldsymbol{P}_i)\). Stochastic selection can be implemented via fitness-proportionate sampling,
\begin{equation}
p_i \;=\; \frac{F_i}{\sum_{j=1}^{N} F_j},
\end{equation}
where \(p_i\) is the probability that individual \(i\) is chosen as a parent, \(F_i\) is its fitness, and \(N\) is the population size.
New offspring are generated by crossover, for example the arithmetic form
\begin{equation}
\boldsymbol{P}^{(\mathrm{child})} \;=\; \lambda\,\boldsymbol{P}^{(\mathrm{parent1})} + (1-\lambda)\,\boldsymbol{P}^{(\mathrm{parent2})},\qquad \lambda\in[0,1],
\end{equation}
where \(\lambda\) is the mixing coefficient weighting the two parents. A mutation step perturbs offspring components,
\begin{equation}
P_k^{(\mathrm{child})} \,\leftarrow\, P_k^{(\mathrm{child})} + \sigma_k\,\xi_k,\qquad \xi_k\sim\mathcal{N}(0,1),
\end{equation}
where \(P_k\) is the \(k\)-th parameter, \(\sigma_k>0\) is the mutation scale, and \(\xi_k\) is standard Gaussian noise. Elitism optionally copies the best individuals unchanged to the next generation, improving stability on rugged landscapes.

\subsubsection*{Particle Swarm Optimization}

Particle Swarm Optimization (PSO) \cite{Kennedy2002ParticleSO} represents each candidate as a particle with position \(\boldsymbol{x}_i^t\) and velocity \(\boldsymbol{v}_i^t\) at iteration \(t\). Let \(\boldsymbol{p}_i\) denote the particle’s personal best and \(\boldsymbol{g}\) the global best in the swarm. The standard update reads
\begin{align}
\boldsymbol{v}_i^{\,t+1} &= \chi\,\boldsymbol{v}_i^{\,t} + c_1\,\boldsymbol{r}_1\odot\big(\boldsymbol{p}_i - \boldsymbol{x}_i^{\,t}\big) + c_2\,\boldsymbol{r}_2\odot\big(\boldsymbol{g} - \boldsymbol{x}_i^{\,t}\big), \label{eq:pso-v}\\
\boldsymbol{x}_i^{\,t+1} &= \boldsymbol{x}_i^{\,t} + \boldsymbol{v}_i^{\,t+1}, \label{eq:pso-x}
\end{align}
where \(\chi\in[0,1]\) is the inertia weight, \(c_1,c_2>0\) are the cognitive and social acceleration coefficients, \(\boldsymbol{r}_1,\boldsymbol{r}_2\sim \mathcal{U}[0,1]^d\) are indipendent identically distributed (i.i.d) vectors of uniform random numbers (drawn componentwise), and \(\odot\) denotes the Hadamard (elementwise) product. Equations \eqref{eq:pso-v}–\eqref{eq:pso-x} bias motion toward the personal and global best positions while retaining momentum through \(\chi\), yielding fast convergence with modest hyperparameter sensitivity.

\subsubsection*{Grey Wolf Optimizer}

Grey Wolf Optimization (GWO) \cite{Mirjalili2014GreyWO} abstracts the encircling and hunting strategy of wolves with a leadership hierarchy. Let \(\boldsymbol{X}_\alpha\), \(\boldsymbol{X}_\beta\), and \(\boldsymbol{X}_\delta\) be the three best solutions at iteration \(t\), and let \(\boldsymbol{X}\) denote the position of a generic wolf. Encircling and guidance are defined by
\begin{align}
\boldsymbol{D}_\alpha &= \big|\boldsymbol{C}_1 \odot \boldsymbol{X}_\alpha - \boldsymbol{X}\big|,\qquad \boldsymbol{X}_1 = \boldsymbol{X}_\alpha - \boldsymbol{A}_1 \odot \boldsymbol{D}_\alpha,\\
\boldsymbol{D}_\beta  &= \big|\boldsymbol{C}_2 \odot \boldsymbol{X}_\beta  - \boldsymbol{X}\big|,\qquad \boldsymbol{X}_2 = \boldsymbol{X}_\beta  - \boldsymbol{A}_2 \odot \boldsymbol{D}_\beta, \\
\boldsymbol{D}_\delta &= \big|\boldsymbol{C}_3 \odot \boldsymbol{X}_\delta - \boldsymbol{X}\big|,\qquad \boldsymbol{X}_3 = \boldsymbol{X}_\delta - \boldsymbol{A}_3 \odot \boldsymbol{D}_\delta,
\end{align}
with the position update
\begin{equation}
\boldsymbol{X}^{\,t+1} \;=\; \frac{\boldsymbol{X}_1 + \boldsymbol{X}_2 + \boldsymbol{X}_3}{3},\qquad 
\boldsymbol{A}_j = 2a\,\boldsymbol{r}_j - a,\quad \boldsymbol{C}_j = 2\,\boldsymbol{r}_j.
\end{equation}
Here \(\boldsymbol{r}_j\sim\mathcal{U}[0,1]^d\), \(\odot\) denotes elementwise multiplication, and \(a\) decreases linearly from \(2\) to \(0\) over iterations. Large \(|\boldsymbol{A}_j|\) encourages exploration early on; as \(a\to 0\) the search contracts around \(\boldsymbol{X}_\alpha,\boldsymbol{X}_\beta,\boldsymbol{X}_\delta\), effecting exploitation.

\subsection{Bayesian Optimization}\label{sec:bayesopt}

Bayesian Optimization (BO) is employed to minimize the black–box objective \(J(\boldsymbol{P})\) in \eqref{eq:cost} under simple box constraints \(\Pi=\{\boldsymbol{P}\in\mathbb{R}^{15}\mid \boldsymbol{\ell}\le\boldsymbol{P}\le\boldsymbol{u}\}\). 
The parameter vector \(\boldsymbol{P}\) stacks five PID triplets \((K_p,K_i,K_d)\).
At each iteration \(t\), BO maintains a probabilistic surrogate of the cost conditioned on the evaluated designs \(\mathcal{D}_t=\{(\boldsymbol{P}_i, J(\boldsymbol{P}_i))\}_{i=1}^t\), and proposes the next candidate by maximizing an acquisition function \(\alpha\) that balances exploration and exploitation:
\[
\boldsymbol{P}_{t+1}\;=\;\arg\max_{\boldsymbol{P}\in\Pi}\;\alpha\big(\boldsymbol{P};\,\mathcal{D}_t\big).
\]
This sequential design enables BO to locate near-optimal solutions in relatively few evaluations, making it particularly well suited when each simulation is expensive. Each function evaluation executes the full composite mission in the simulator (aerodynamics, noise model, and, when enabled, Dryden turbulence), returning the scalar cost \(J(\boldsymbol{P})\). 
This setup exploits BO’s sample efficiency when each rollout is relatively expensive, while keeping the dimensionality moderate so that surrogate modeling remains effective.

\subsection{Reinforcement Learning}\label{sec:rl_onestep}

We conceptualized our RL implementation with a question: can a DRL agent capture the essence of an optimization problem when framed with the most elementary action–return formulation? Our curiosity emerged from the possibility that modern DRL algorithms could serve as general-purpose optimizers for tasks that cannot be easily broken down into multi-step episodes with structured intermediate rewards. This perspective is especially relevant for classes of control and design problems, such as PID controller tuning or aerodynamic shape optimization, where the evaluation of a candidate solution often provides only a single terminal reward without meaningful intermediate feedback.

In this context, we formulated PID tuning as a continuous, bound--constrained one step MDP. There is no notion of state transitions or long-term planning within an episode: each episode consists of a single decision and immediate terminal feedback. In the other hand, a fully multi-state RL formulation would treat the PID gains as adaptive variables that evolve with the flight state and environment, yielding a dynamic, context-dependent controller, i.e. adaptive control. While such an approach is promising, it is beyond our current scope and strays from previous considerations. Consequently, the problem reduces to learning a sampling policy that maximizes expected immediate reward, without modeling state transitions or intra-episode adaptation. 
Framing the problem in this way allows us to investigate whether general-purpose RL algorithms can effectively solve static parameter selection under noisy returns, and to directly compare the learned sampling policy against traditional optimization methods without the need to redesign the control architecture.

Thus, the MDP schema is modeled as follows. 
Let $\boldsymbol{P}\in\Pi\subset\mathbb{R}^{15}$ denote the vector of tunable PID gains for the six optimized loops: $x$ and $y$ position, altitude, roll--pitch attitude, horizontal speed, and vertical speed, while yaw gains are kept fixed to reduce dimensionality. 
The feasible set is the hyper-rectangle
\begin{equation}
\Pi \;=\; \bigl\{\, \boldsymbol{P}\in\mathbb{R}^{15} \;\big|\; \boldsymbol{\ell}\le \boldsymbol{P}\le \boldsymbol{u} \,\bigr\},
\end{equation}
with bounds $\boldsymbol{\ell},\boldsymbol{u}\in\mathbb{R}^{15}$ chosen consistently with the ranges used both in the metaheuristics and BO. 

An action is the gain vector $a\equiv\boldsymbol{P}\in\Pi$, which is decoded into grouped $(K_p,K_i,K_d)$ triplets for the six loops. 
Executing $a$ triggers a closed-loop simulation on the composite mission with those gains and returns the scalar objective $J(\boldsymbol{P})$ defined by the cost function. 
The (undiscounted) reward is
\begin{equation}
r(\boldsymbol{P}) \;=\; -\,J(\boldsymbol{P}) \, .
\end{equation}

Each episode has horizon $H=1$: the agent proposes $\boldsymbol{P}$, the simulator returns $r(\boldsymbol{P})$, and the episode terminates. 
The observation (context) $s$ is a bounded vector in $\mathbb{R}^{15}$ used only to condition the policy. 
In practice, it is either fixed to a nominal initialization or independently resampled at reset from a prescribed distribution. 
Since episodes terminate immediately, the transition kernel is trivial and the value function reduces to expected immediate reward. 
The learning objective is therefore
\begin{equation}
\max_{\pi}\;\; \mathbb{E}_{\,s\sim\rho,\;\boldsymbol{P}\sim\pi(\cdot\mid s)}\bigl[\,r(\boldsymbol{P})\,\bigr]
\;=\; -\,\min_{\pi}\;\mathbb{E}_{\,s\sim\rho,\;\boldsymbol{P}\sim\pi(\cdot\mid s)}\bigl[\,J(\boldsymbol{P})\,\bigr],
\end{equation}
where $\rho$ denotes the reset distribution over contexts. 

Optional stochasticity is introduced by enabling Dryden turbulence and waypoint randomization during rollouts, so the return is noisy even for fixed $\boldsymbol{P}$. 
From the agent’s standpoint, this corresponds to i.i.d. bandit pulls with stochastic rewards, provided resets and seeds are independent across episodes~\cite{silver2021reward}.

We employ off--policy actor--critic methods with replay to learn a proposal (sampling) policy over $\Pi$. 
Although such algorithms are generally designed for multi-step control, in the single-step setting their Bellman targets collapse to the immediate reward, so they act as entropy-regularized derivative-free optimizers with learned proposal distributions.

\subsubsection*{Soft Actor--Critic}
Soft Actor--Critic (SAC) is an off--policy method that normally learns a stochastic policy by maximizing expected return together with an entropy bonus that promotes exploration~\cite{Haarnoja2018SoftAO}. 
In our one--step formulation, the agent proposes a gain vector $\boldsymbol{P}$, receives an immediate reward $r(\boldsymbol{P})$, and the episode terminates. 
Hence, the usual soft Bellman recursion collapses to a direct target
\begin{equation}
y \;=\; r(\boldsymbol{P}) ,
\end{equation}
so each critic $Q_{\psi_j}$ is simply trained to regress onto the observed reward:
\begin{equation}
\mathbb{E}\!\left[\bigl(Q_{\psi_j}(s,\boldsymbol{P}) - y\bigr)^2\right], \qquad j\in\{1,2\}.
\end{equation}
The actor update retains its entropy--regularized form,
\begin{equation}
J_\pi(\phi) \;=\; \mathbb{E}_{\,s\sim\mathcal{D},\;\boldsymbol{P}\sim\pi_\phi(\cdot\mid s)}
\!\left[\,\alpha \log \pi_\phi(\boldsymbol{P}\mid s) \;-\; \min_{j} Q_{\psi_j}(s,\boldsymbol{P})\,\right],
\end{equation}
where the temperature $\alpha$ is adapted to match a target entropy. 
Actions are rescaled to lie inside $\Pi$. 
Because there are no future rewards, SAC here learns a probability distribution over $\Pi$ that places higher density on low--cost regions while maintaining randomness to keep exploration active.

\subsubsection*{Twin Delayed DDPG}
Twin Delayed DDPG (TD3) is an off--policy algorithm that learns a deterministic actor $\mu_\phi(s)$ 
with two critics to counteract value overestimation~\cite{Dankwa2019TwinDelayedDA}. 
In a standard multi--step setting, targets would include bootstrapped estimates of future value. 
Here, with horizon one, the target is purely the observed reward:
\begin{equation}
y \;=\; r(\boldsymbol{P}), \qquad 
\boldsymbol{P}=\mu_{\bar{\phi}}(s)+\varepsilon,\;\; 
\varepsilon\sim\mathcal{N}(0,\sigma^2)\;\text{(clipped)},
\end{equation}
where small Gaussian noise is added for target policy smoothing. 
Each critic is trained by minimizing
\begin{equation}
\mathbb{E}\!\left[\bigl(Q_{\psi_j}(s,\boldsymbol{P}) - y\bigr)^2\right],
\end{equation}
and the actor parameters are updated less frequently using the deterministic policy gradient
\[
\nabla_\phi J(\phi) \;\approx\; \nabla_\phi Q_{\psi_1}\bigl(s,\mu_\phi(s)\bigr).
\]
All proposed actions are projected into $\Pi$. 
The update delay and smoothing mechanisms stabilize training even in this simplified one--step setting.

\subsection{Considerations}\label{sec:discussion}

To ensure the optimization loop remains computationally feasible while still capturing the main aerodynamic and acoustic mechanisms, we adopt a layered modeling strategy. Aerodynamics are evaluated through a surrogate trained on BEMT to avoid costly high-fidelity sweeps, and acoustic emissions are estimated per rotor in 1/3-octave bands, propagated on an $N\times N$ ground grid with radius-based culling so that only cells above a relevant threshold are updated. In practice, $N$ is chosen so that SPL maps change by less than about 1 dB upon refinement, while the active radius is derived online from instantaneous source level, spherical spreading, and optional ISO 9613-1 absorption to skip far-field cells with negligible contribution. This design keeps computational load manageable and enables the large number of rollouts required by optimization algorithms. 

Within the optimization loop, the cost function is defined using proxies that balance accuracy and efficiency. For acoustic performance, the cost uses a SWL-based measure that combines a high-order norm of the sound power level time series with its maximum value, penalizing both sustained energy and peaks. This formulation captures the main characteristics of rotor noise while remaining computationally inexpensive, but neglects detailed frequency-domain features such as tonal harmonics and fine-grained modulation effects, which may be relevant for high-fidelity acoustic characterization, or psychoacoustic evaluations. Similarly, the aerodynamic surrogate abstracts away wake interactions and coherence effects between rotors. However, for controller optimization the chosen formulation is adequate, as it enables thousands of rollouts with consistent ranking of candidate controllers, while more detailed models would render the loop intractable.  

A final point concerns the RL formulation. In standard multi-step settings, SAC and TD3 rely on bootstrapping to propagate value estimates across time, but in our one-step formulation each trial terminates immediately, so their Bellman targets collapse to the observed reward. This makes the critics essentially supervised regressors of $r(\boldsymbol{P})$, while the actors shape how new candidate gains are sampled. SAC maintains a stochastic policy with an entropy bonus, so in this context it acts as a learned probability distribution that gradually concentrates mass on promising regions while still exploring alternatives. TD3 instead maintains a deterministic actor corrected by noise injection and delayed updates, effectively functioning as a learned point estimator refined by controlled perturbations. 
Both algorithms can therefore be understood as adaptive proposal mechanisms: SAC as a stochastic sampler and TD3 as a deterministic optimizer with smoothing. 
From a practical perspective, this places them conceptually close to classical optimizers: SAC behaving like an adaptive random search biased toward low-cost regions, and TD3 like a gradient-driven descent with jitter added for robustness; yet with the distinction that their sampling strategies are learned online from experience rather than hand-crafted. In our setting, these methods are not expected to yield the most efficient optimization compared to specialized algorithms; rather, they are included to examine how general-purpose RL behaves when cast as a static parameter selection problem with noisy rewards, and to assess whether their adaptive sampling provides any advantage in a one-step formulation.


\section{Results}\label{sec:results}

This section presents how the optimization campaign was set up. The section first introduces the simulation setup and test case, then reports results under progressively constrained conditions, and finally discusses the computational overhead and the performance of the optimized controllers on unseen missions.

\subsection{Test Cases}

The optimization framework was evaluated using a simulated quadrotor model based on the DJI Matrice~300 platform, characterized by a mass of 5.2 kg and arm length of 0.32 m. Controller-imposed speed constraints limited maximum horizontal and vertical velocities to \(13.89~\mathrm{m/s}\) and \(5.56~\mathrm{m/s}\), respectively. Each simulation ran for a fixed mission duration of 150 seconds with a sampling frequency of 125 Hz (0.008-second intervals), resulting in approximately 15.0 seconds of wall clock computation time on our workstation with Intel Xeon W-2135 3.70Ghz, 6 core, 12 threads CPU.

All optimization algorithms were tested under varying initial conditions, specifically by changing the PID search bounds (maximum and minimum explorable values), by enabling or disabling wind turbulence simulation, and by using an initial PID set obtained through manual tuning. We present the results in three steps:
\begin{enumerate}
    \item We start by using wide bounds for the search space of gains value. This allows us to test each method's ability to explore the space without an initial guess.
    \item Then, based on manual tuning, we restrict bounds, and we use warm start on manual tuned gains themselves, without turbulence.
    \item Third, the same setting as second step, with Dryden turbulence for evaluating robustness and generalization.
\end{enumerate}

In the final discussion, we compare the computation time per iteration and analyze the best controller for an unseen mission in terms of trajectory, attitude, noise, power, and cost decomposition.

The manual tuning followed the Ziegler–Nichols procedure \cite{meshram2012tuning}: for each loop under test we set \(K_i=0\) and \(K_d=0\), applied a step in the corresponding reference (hover-to-altitude step for altitude; small position hops for horizontal motion), and increased \(K_p\) until a sustained oscillation with nearly constant amplitude was observed. The associated ultimate gain \(K_u\) and ultimate period \(T_u\) were then mapped to PID gains via the Ziegler–Nichols formulas \(K_p = 0.6\,K_u\), \(T_i = T_u/2\) (so \(K_i = K_p/T_i\)), and \(T_d = T_u/8\) (so \(K_d = K_p T_d\)). This baseline produced a total cost \(J=218.2\) in the composite mission with no turbulence, and is used as a reference line in the following analyses. 

\subsection{Optimization performance}

\paragraph{\textbf{(1) No initial guess/large bounds}} The algorithms were initially tested with a fixed evaluation budget of 6{,}000 steps, using wide parameter bounds and no initial guesses. 
The aim was not to outperform manual tuning, but rather to examine the extent to which each method could explore the search space and identify viable gain regions without any prior information. All PID search bounds were set uniformly in the range \([0,\,1\times 10^{5}]\). Since this range defines an extremely large search space, we cannot expect better results compared to manual tuning.

\begin{figure}[H] 
    \centering 
    \includegraphics[width=1\linewidth]{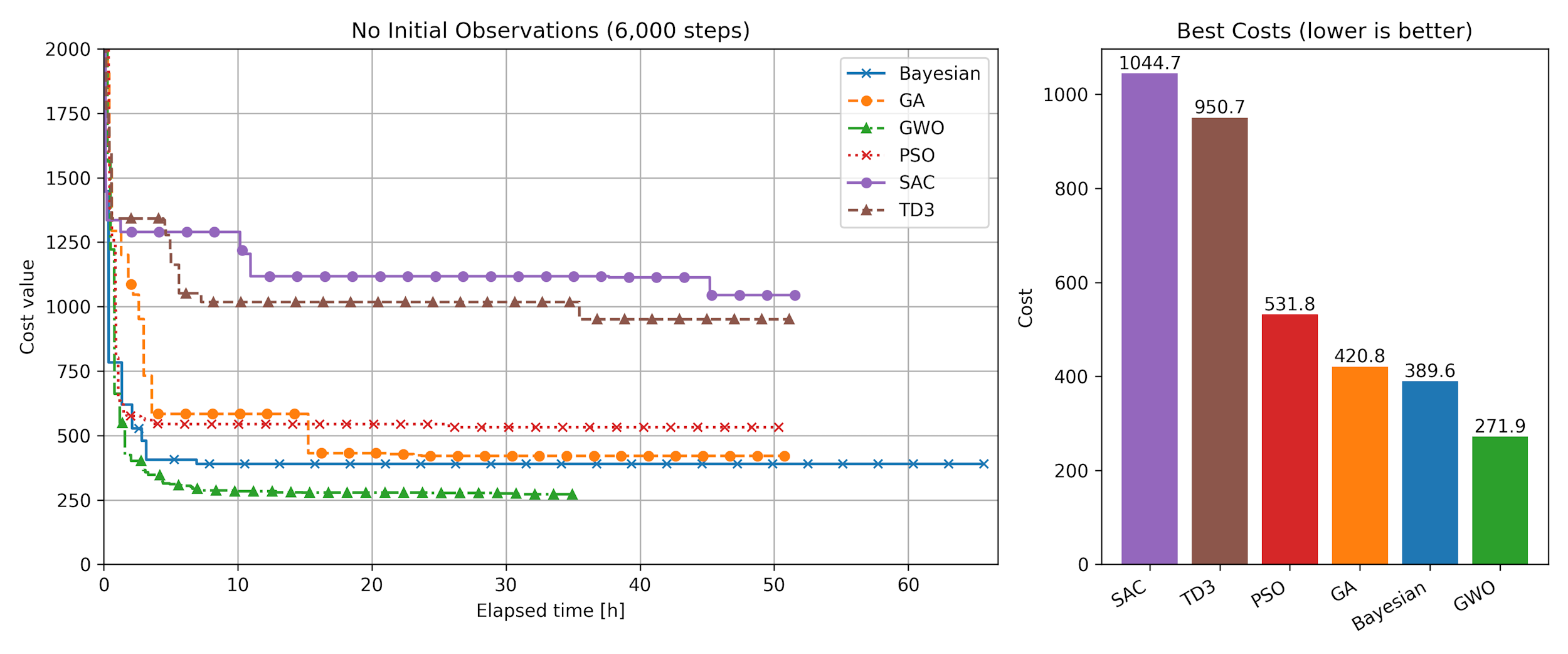} 
    \caption{Wide–bound, no–warm–start study (\(6{,}000\) evaluations). Left: minimum cost versus wall time; Right: best cost achieved at the step budget limit.} \label{fig:exp1} 
\end{figure} 

Figure~\ref{fig:exp1} illustrates that metaheuristic methods and BO outperform DRL methods, both in terms of convergence speed and final cost reduction. 

The left panel reports the temporal evolution of the minimum observed cost, while the right panel presents a bar chart comparing the best final results after 6{,}000 evaluations. Among the tested algorithms, GWO achieved the best performance, converging to a final cost of \(271.9\), which remains slightly higher than the baseline obtained from manual Ziegler--Nichols tuning. BO ranked second, followed by GA and PSO.

DRL methods, in contrast, exhibited slower progress and higher variance across runs. In this configuration the algorithms spent a significant portion of the budget maintaining exploration rather than exploiting promising regions, which hindered convergence within the 6{,}000-step limit. Under such conditions, RL can identify feasible regions but struggles to match the efficiency of specialized metaheuristics or BO.

\paragraph{\textbf{(2) Small bounds, warm start}} After restricting the bounds based on information from manual tuning, a second set of experiments was performed with a fixed wall–time budget of 14~hours. The aim was to assess the extent to which optimization methods could improve the performance of the manually tuned PID configuration. The search space bounds were defined heuristically by inspecting the sensitivity of each parameter to overshoot and oscillatory behavior during the manual tuning. In addition, all algorithms were initialized with an initial observation corresponding to the specific set of PID gains obtained with manual tuning (warm start). For DRL methods, the first action was forced to evaluate this initial set. Optimizations were conducted first in a disturbance–free environment and then with Dryden turbulence enabled.

\begin{figure}[H]
    \centering
    \includegraphics[width=1\linewidth]{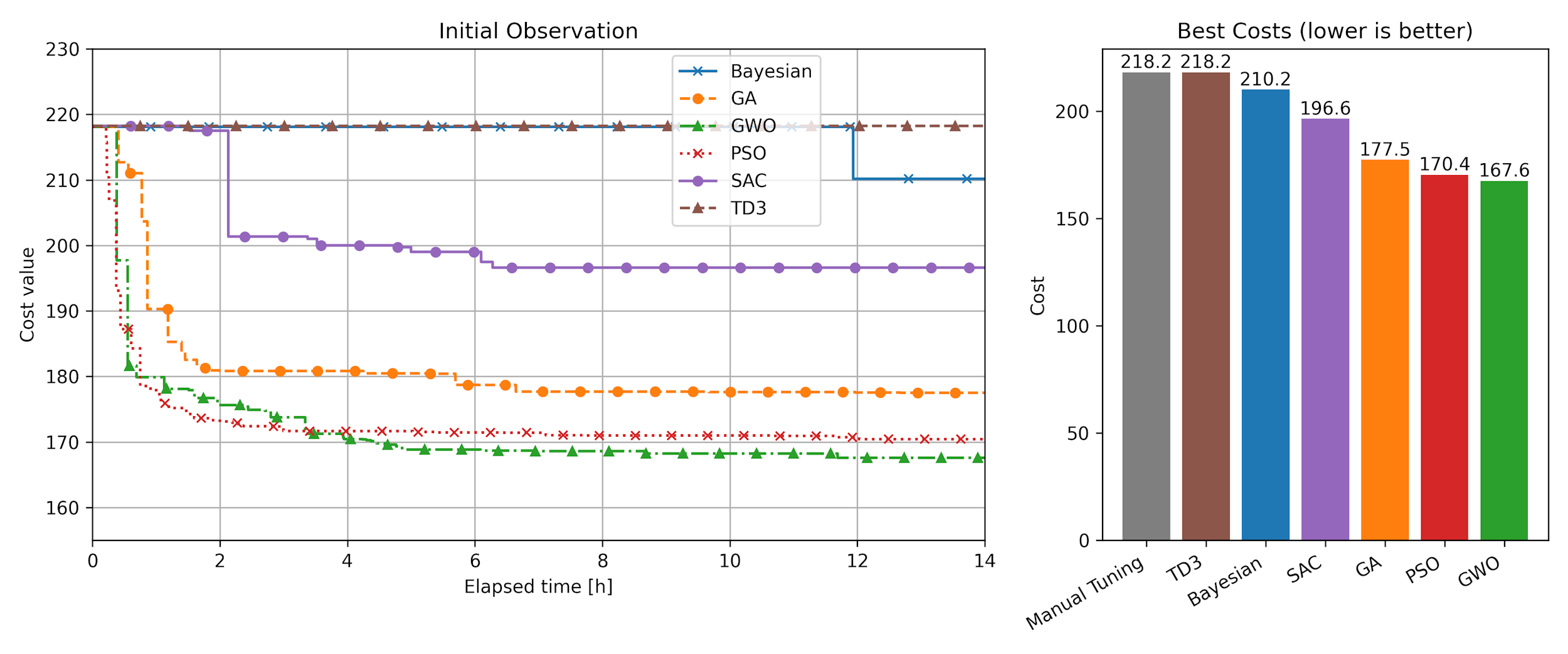}
    \caption{Bounded search with warm start, fixed 14-hour budget, no turbulence. Left: incumbent cost versus wall time; Right: best cost at time budget limit.}
    \label{fig:exp2}
\end{figure}

Figure~\ref{fig:exp2} shows the results for the bounded search with warm start and no turbulence. All metaheuristic algorithms were able to improve upon the manual baseline within the given time budget, with GWO again achieving the lowest final cost, followed by PSO, GA, and BO. RL, in the one-step-like setup adopted here, did not yield noticeable improvement over the baseline. While the GWO achieved the best performance in this study, reducing the cost by about 23$\%$ compared to manual tuning, this does not imply it is universally superior: metaheuristic performance is inherently problem-dependent and varies with mission scenarios or cost formulations \cite{wolpert2002no}.

\begin{figure}[H]
    \centering
    \includegraphics[width=1\linewidth]{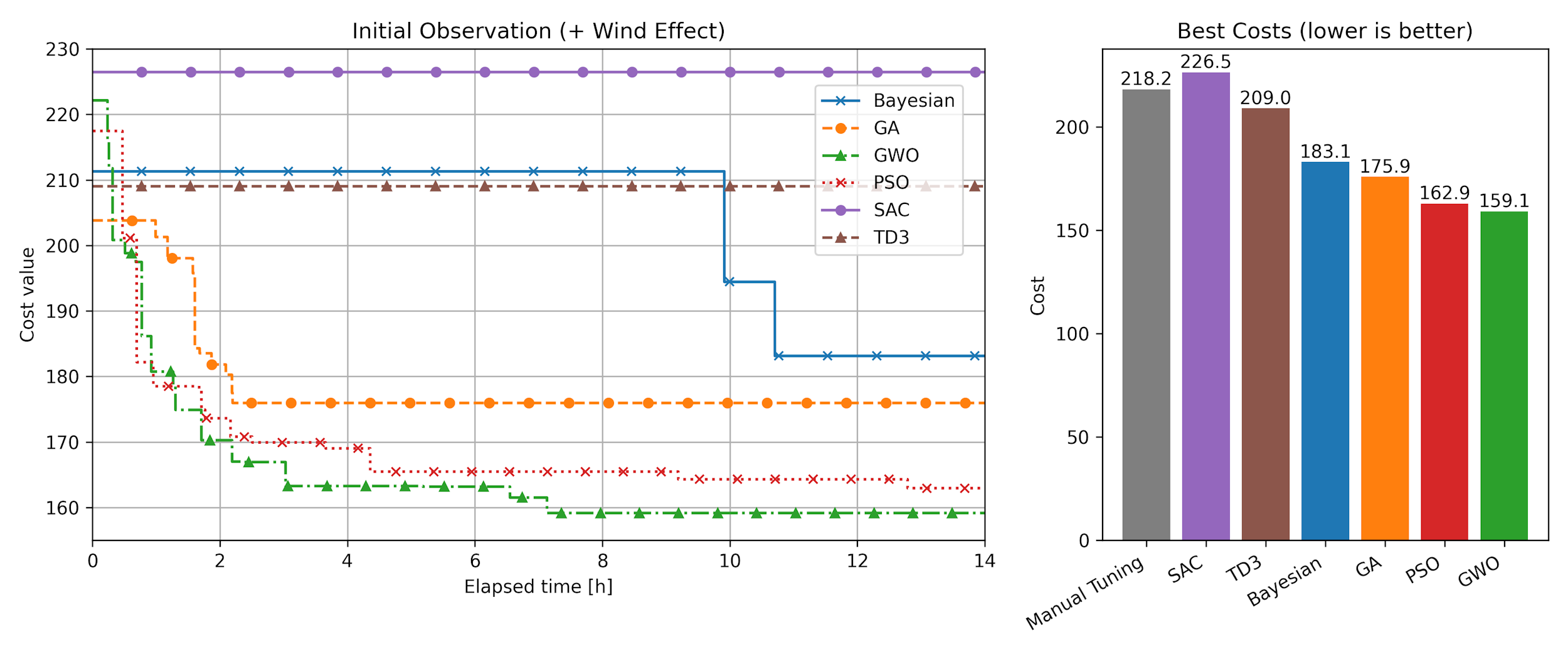}
    \caption{Bounded search with warm start, fixed 14-hour budget, with Dryden turbulence enabled. } 
    \label{fig:exp3}
\end{figure}

\paragraph{\textbf{(3) Small bounds, warm start with wind turbulence}} 
When Dryden turbulence is enabled, the additional disturbances increase the ability of the optimizers to discriminate between candidate solutions, leading to more robust and generalizable tuning outcomes (Figure~\ref{fig:exp3}). 
GWO again delivered the best performance, reducing the cost by 27\% relative to the manual baseline, with PSO and GA achieving similar but slightly weaker results. 
BO converged more slowly, while DRL methods continued to show no measurable improvement. 
These findings indicate that, in a one–shot action–reward setting, DRL is poorly suited for static PID gain optimization.  

Overall, these experiments confirm that optimization algorithms can substantially improve PID tuning compared to manual heuristics. 
Among the tested approaches, GWO consistently achieved the best performance, while BO proved effective at exploring wide parameter ranges but was less successful in fine-tuning around specific regions.
In contrast, DRL under the one-step formulation showed limited effectiveness, suggesting that such methods may be better suited to multi-step, dynamic control tasks, such as learning adaptive or state-dependent PID policies, rather than static parameter selection.

\subsection{Discussion}

This section provides an interpretation of the outcomes of the optimization campaign. The optimal cost values obtained through the different optimization algorithms are summarized in Table \ref{tab:performance_table}. In particular,Test Case 1 refers to a wide parameter search without an initial guess, Test Case 2 corresponds to a bounded search with a warm start in the absence of turbulence, and Test Case 3 denotes a bounded search with a warm start under Dryden turbulence conditions. Then, we quantified the additional computation introduced by each method relative to the baseline simulator, assessed the best configuration for a different mission, and analyzed the improvements compared to manual tuning.

\begin{table}[H]
    \centering
    \small
    \caption{Optimal costs obtained with different optimization algorithms across the three test cases. Lower values indicate better performance.}

    \label{tab:performance_table}
    \renewcommand{\arraystretch}{1.2}
    \begin{tabular}{l P{1.5cm} P{1.5cm} P{1.5cm} P{1.5cm} P{1.5cm} P{1.5cm}}
    \hline
    \textbf{} & \textbf{GWO} & \textbf{PSO} & \textbf{GA} & \textbf{SAC} & \textbf{TD3} & \textbf{BO} \\ \hline
    Test Case 1 & 271.9 & 531.8 & 420.8 & 1044.7 & 950.7 & 389.6\\
    Test Case 2 & 167.6 & 170.4 & 177.5 & 196.6 & 218.2 & 210.2\\
    Test Case 3 & \textbf{159.1} & 162.9 & 175.9 & 226.5 & 209.0 & 183.1 \\ \hline
    \end{tabular}
\end{table}

\subsubsection*{Computational overhead}

Table~\ref{tab:avg_time} reports the average computation time per optimization step for all tested algorithms during the fixed wall-time experiments. This is to reduce the occurrences of short simulations that are interrupted prematurely due to highly inaccurate PID configuration proposals. Results show that metaheuristic methods exhibit nearly identical runtimes, with GWO, PSO, and GA requiring approximately \(17\)~s per iteration. Reinforcement Learning methods, SAC and TD3, are slightly slower, with TD3 in particular showing a noticeable increase in computational demand. BO resulted in the highest runtime, averaging more than \(34\)~s per iteration, due to the additional cost of surrogate model fitting and acquisition function maximization at each step.

\begin{table}[H]
\centering
\small
\caption{Average computation time per iteration for all evaluated optimization algorithms. Algorithm overhead represents the additional computational cost imposed by each optimization method beyond the baseline single-step simulation time of 15.0 seconds.}
\label{tab:avg_time}
\renewcommand{\arraystretch}{1.2}
\begin{tabular}{l P{3cm} P{3cm}}
\hline
\textbf{Algorithm} & \textbf{Average time per iteration [s]} & \textbf{Algorithm Overhead [\%]} \\ \hline
GWO & 17.04 & 13.6\\
PSO & 17.37 & 15.8 \\
GA & 17.66 & 17.7 \\
SAC & 17.95 & 19.6 \\
TD3 & 22.56 & 50.4 \\
BO & 34.39 & 102.6 \\ \hline
\end{tabular}
\end{table}

\subsubsection*{Performance analysis} 

\begin{figure}[H]
    \centering
    \includegraphics[width=1\linewidth]{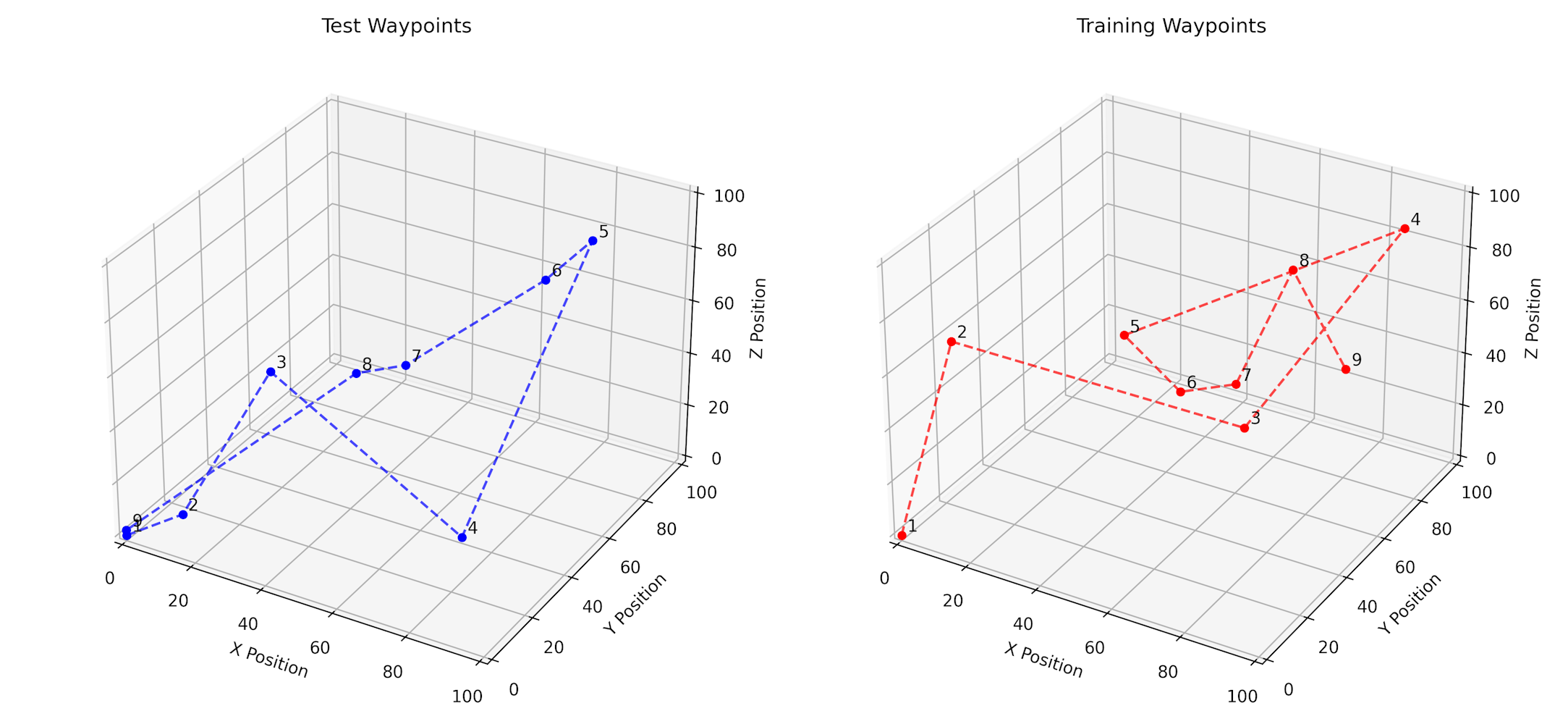}
    \caption{Comparison between the training and testing waypoint trajectories used for evaluating generalization of the optimized controllers.}
    \label{fig:wp_comp}
\end{figure}

The final stage of the analysis evaluates the performance of the optimized controller on relevant cost components, comparing the PID configuration obtained with GWO under turbulent conditions against the manually tuned baseline. To assess generalization, this evaluation was performed on a mission trajectory different from the one used during training, thereby testing whether the optimized gains transfer effectively to new flight conditions rather than overfitting to the composite path employed during optimization. The training and test trajectories are illustrated in Figure~\ref{fig:wp_comp}.

Before turning to cost decomposition, we first compare the main dynamic signals to illustrate how optimization affects stability, noise, and power. These qualitative trends provide context for the quantitative improvements reported later.

\begin{figure}[H]
    \centering
    \includegraphics[width=1\linewidth]{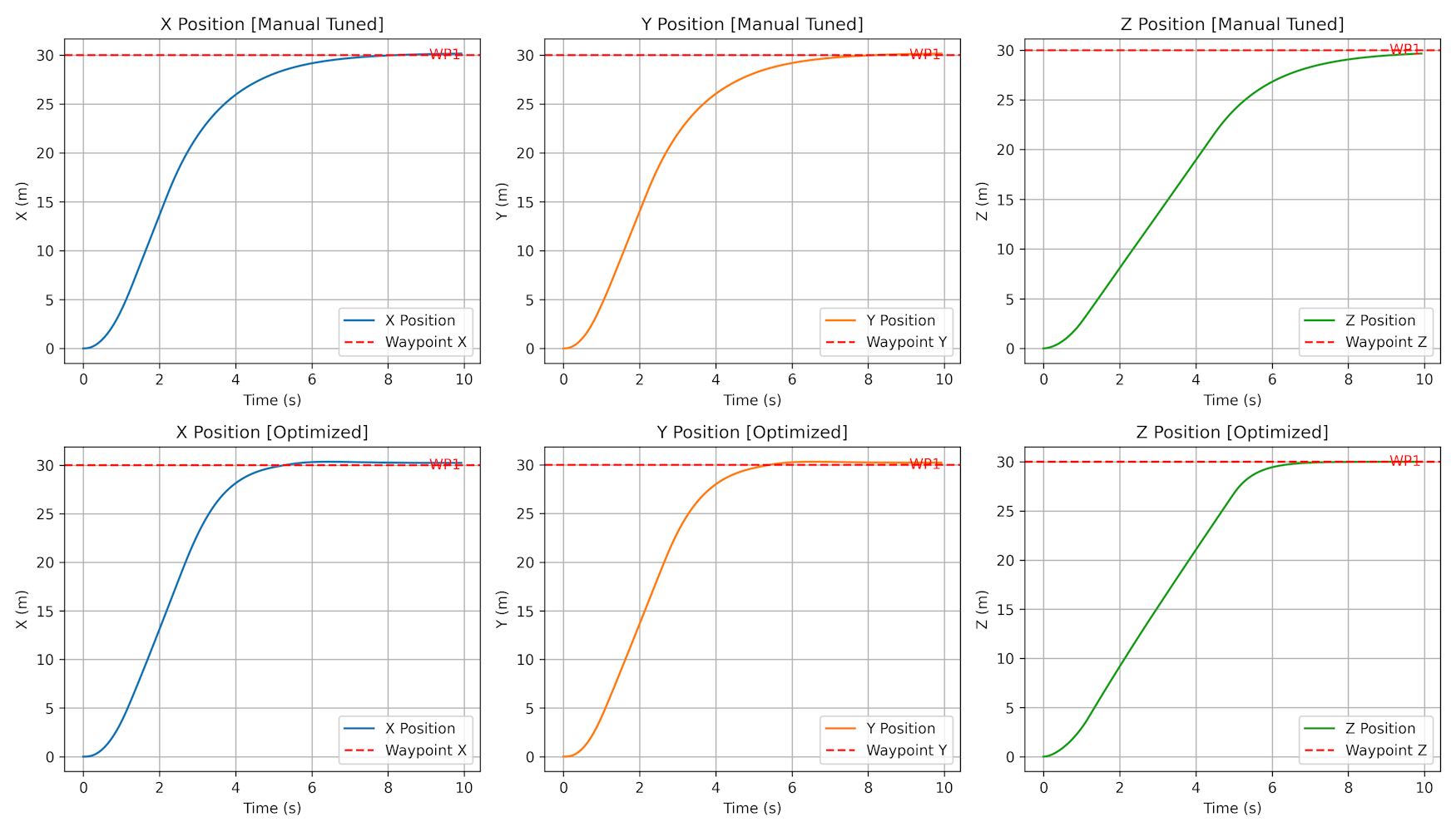}
    \caption{Trajectory tracking during the test mission. 
    Top row: manual baseline (Ziegler--Nichols). 
    Bottom row: GWO-optimized.}
    \label{fig:trajectory}
\end{figure}

As shown in Figure~\ref{fig:trajectory}, the optimized controller converges more rapidly to the reference trajectory. In particular, GWO approach completes the mission in 55.79 seconds, compared to 64.62 seconds obtained with the manually tuned controller. 
For clarity, the figure illustrates a single target-reaching maneuver rather than the entire mission. Beyond this faster convergence, no substantial trajectory improvements are observed, since the Ziegler–Nichols method already yields satisfactory stabilization of the spatial response.
The main benefits of optimization instead emerge in the indirect objectives included in the cost function, reduced oscillations, lower acoustic emissions, and improved power efficiency.

\begin{figure}[H]
    \centering
    \includegraphics[width=1\linewidth]{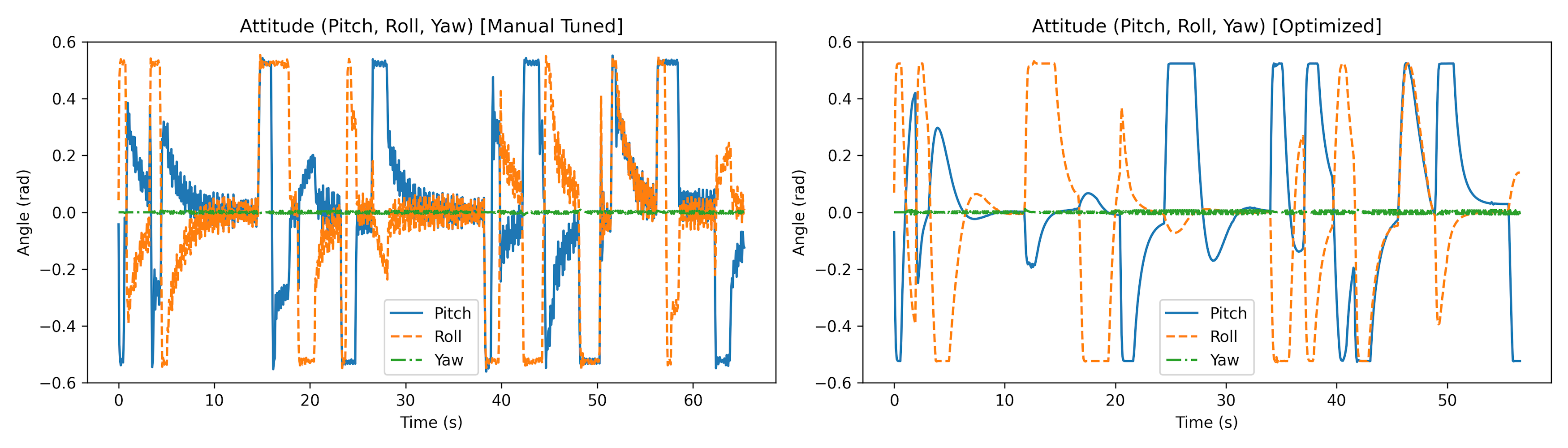}
    \caption{Time histories of pitch, roll, and yaw during the test mission. Left: manual baseline. Right: GWO-optimized.}
    \label{fig:pitch_roll_yaw}
\end{figure}

The evolution of attitude angles is shown in Figure~\ref{fig:pitch_roll_yaw}. Compared with the manually tuned baseline, the GWO-optimized controller yields visibly smoother profiles with reduced oscillatory behaviour across pitch, roll, and yaw. This indicates that the optimization substantially improves closed-loop stability, with the controller avoiding unnecessary corrective actions that typically amplify oscillations and contribute to both excess power consumption and tonal noise generation.

\begin{figure}[H]
    \centering
    \includegraphics[width=1\linewidth]{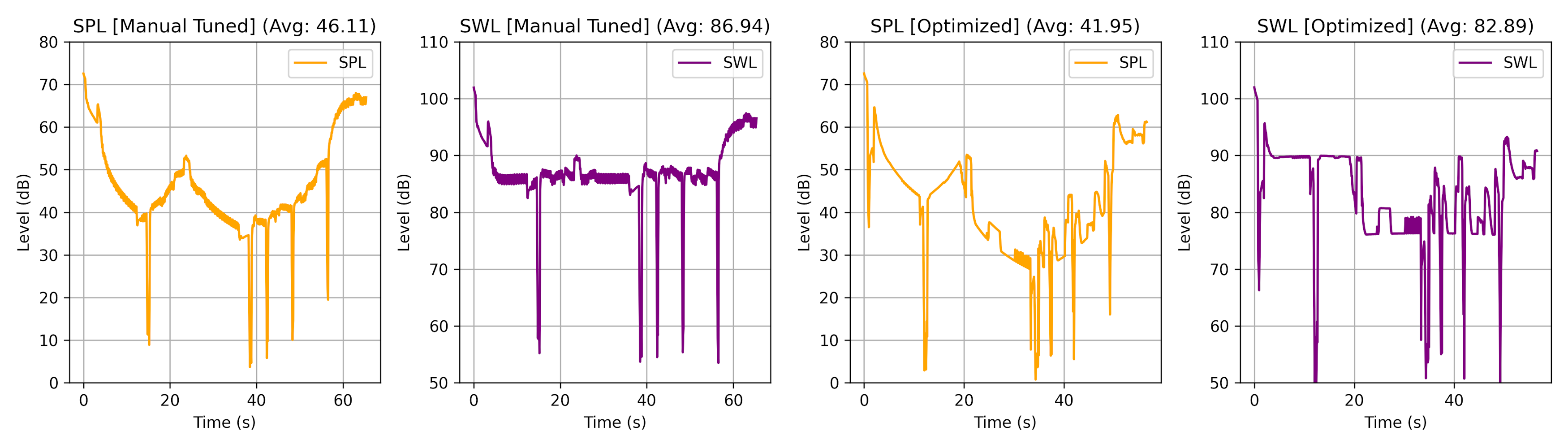}
    \caption{Comparison of SPL (receivers) and SWL (source) time series. First and second images: manual baseline. Third and forth images: GWO-optimized.}
    \label{fig:sound}
\end{figure}

The acoustic results are presented in Figure~\ref{fig:sound}, which compares broadband averaged SPL at receivers and SWL at the source. Averaged SPL is calculated as the mean for each cell area for each timestep. In both cases, the optimized controller consistently achieves lower acoustic levels, confirming that the improvements in stability and smoother thrust commands translate directly into a reduced acoustic footprint. Importantly, the reductions are not transient but persist throughout the mission, which suggests that the optimization improves the overall noise profile rather than isolated peaks.

\begin{figure}[H]
    \centering
    \includegraphics[width=1\linewidth]{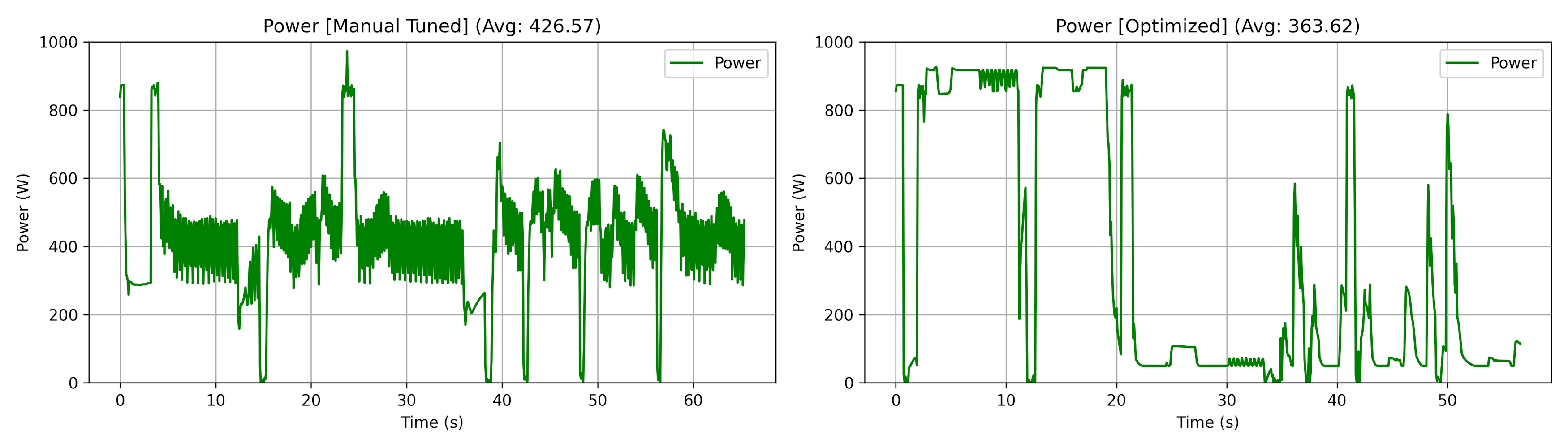}
    \caption{Time evolution of electrical power demand. Left: manual baseline. Right: GWO-optimized.}
    \label{fig:power}
\end{figure}

Figure~\ref{fig:power} reports the electrical power demand over time. The optimized controller exhibits smoother consumption with fewer fluctuations, leading to an average reduction in power usage. This not only reflects greater efficiency in trajectory execution but also indicates reduced stress on the propulsion system, which could translate into improved endurance and component lifetime in real-world operations.

The quantitative breakdown in Table~\ref{tab:ablations} highlights where the most significant improvements occur. The GWO-optimized configuration reduces the overall cost by 42.7\%, with the largest relative improvements obtained in attitude oscillations (\(-77.8\%\)), noise emissions (\(-35.9\%\)), and power consumption (\(-25.7\%\)). Mission time is also improved (\(-13.6\%\)), while overshoot remains essentially unchanged, indicating that the optimization focuses primarily on stabilizing dynamics and improving efficiency without sacrificing trajectory accuracy. These results underline the limitations of manual tuning: although it can yield acceptable trajectory tracking, it does not provide systematic control over secondary objectives such as energy usage or acoustic footprint, which are critical in urban UAV operations.

\begin{table}[H]
\centering
\small
\caption{Cost decomposition on the unseen mission: manual baseline versus GWO-optimized controller (with turbulence). Percentage change is computed as \((\text{baseline}-\text{optimized})/\text{baseline}\).}
\label{tab:ablations}
\renewcommand{\arraystretch}{1.2}
\begin{tabular}{lrrr}
\hline
\textbf{Term} & \textbf{Baseline} & \textbf{Optimized} & \textbf{Change} \\
\hline
Total cost                    & 289.59 & 165.93 & \(-42.7\%\) \\
Mission time cost             &  64.62 &  55.79 & \(-13.6\%\) \\
Attitude oscillation cost     & 119.23 &  26.47 & \(-77.8\%\) \\
Overshoot cost                &  35.34 &  34.46 & \(-2.5\%\)  \\
Power cost                    &  34.52 &  25.65 & \(-25.7\%\) \\
Noise proxy cost (SWL-based)  &  34.03 &  21.79 & \(-35.9\%\) \\
\hline
\end{tabular}
\end{table}

Complementary indicators are reported in Table~\ref{tab:acoust_energy}. On average, SPL at receivers decreases by 4.16~dB, while SWL at the source is reduced by 4.05~dB. In parallel, average electrical power drops by 14.8\%. These reductions confirm that the benefits of optimization are not limited to abstract cost terms but also manifest as tangible improvements in acoustic footprint and energy efficiency. In the psychoacoustics literature, reductions in the order of 3--5~dB are typically regarded as perceptually significant for human listeners, especially in outdoor urban contexts where background levels are variable but relatively low~\cite{Fastl2007PsychoacousticsSQ, zwicker2013psychoacoustics}. This indicates that the optimized controllers not only improve technical performance but also achieve noise reductions that are likely to be noticeable to end users and communities. 

\begin{table}[H]
\centering
\small
\caption{Acoustic and energy indicators on the unseen mission (arithmetic means over time).}
\label{tab:acoust_energy}
\renewcommand{\arraystretch}{1.2}
\begin{tabular}{lrrr}
\hline
\textbf{Indicator} & \textbf{Baseline} & \textbf{Optimized} & \textbf{Change} \\
\hline
Avg SPL at receivers [dB] & 46.11 & 41.95 & \(-4.16~\mathrm{dB}\) \\
Avg source SWL [dB]       & 86.94 & 82.89 & \(-4.05~\mathrm{dB}\) \\
Avg electrical power [W]  & 426.57 & 363.62 & \(-14.8\%\) \\
\hline
\end{tabular}
\end{table}


\section{Conclusions}\label{sec:concl}

This study presented a systematic comparison of metaheuristic algorithms, BO, and RL for noise-aware PID tuning of a quadrotor UAV within a unified simulation framework. Among the tested approaches, metaheuristics, particularly GWO, consistently provided the most effective and robust improvements under fixed computational budgets, yielding significant reductions in cost and noise-related terms. BO achieved competitive performance but required substantially higher evaluation times per iteration. By contrast, in the present one-step formulation, SAC and TD3 did not outperform the manually tuned baseline, suggesting that RL in its static form is ill-suited for this problem. Nevertheless, the ability of RL to adapt policies over time makes it a promising direction if multi-step episodes or phase-dependent gain scheduling are introduced. Also, controllers tuned with the proposed pipeline generalized well to unseen missions, indicating that the use of a composite training trajectory and turbulence randomization successfully promoted robust solutions that extend beyond the specific optimization scenario.  
The test simulation demonstrates that optimization produces controllers that not only generalize to unseen scenarios but also deliver meaningful practical benefits: smoother dynamics, lower noise, and improved efficiency. These improvements directly address key barriers to UAV adoption in urban contexts, where both acoustic acceptability and energy constraints are critical.
Despite the promising achieved results, a number of challenges remain, opening directions for future work. First, noise optimization was constrained to simplified proxies based on SWL history and peaks. A more sophisticated pipeline including audio-rate auralization and compliant psychoacoustic models would allow direct integration of perceptual metrics into the optimization loop, potentially improving noise-aware control. Second, only PID gain values were optimized, while anti-windup limits, command saturations, and yaw gains remained fixed; expanding the search space to include these parameters could further improve performance. Third, the aerodynamic surrogate relied on a BEMT-generated dataset; extending the surrogate with higher-fidelity data, or applying data fusion and transfer-learning methods would enhance physical fidelity without compromising efficiency \cite{vaiuso2025multi}. Finally, RL approaches could be reformulated to exploit their full potential: instead of treating the problem as a one-step action, where the action provides the complete PID vector, multi-phase formulations could assign gain sets to flight phases (e.g., takeoff, cruise, landing), with state summaries including error norms, wind estimates, and saturation time, and rewards shaped by phase-wise trajectory tracking, control effort, and SPL.


\section*{Acknowledgments}
This research was funded by the Swiss Federal Office of Civil Aviation (FOCA) under the Spezialfinanzierung Luftverkehr programme, measure SFLV 2022-047, "DRACONIAN: DRone trAjectory and CONtrol optImisAtioN for noise emissions reduction".

\appendix

\section{BEMT}\label{appx:bemt}
From momentum theory, for a hovering rotor of radius $R$, the thrust $T_{\Sigma}$ is related to the uniform induced velocity at the disk $v_{\mathrm{ind},0}$ by
\begin{equation}
T_{\Sigma} = 2 \rho A v_{\mathrm{ind},0} \left( V_\infty + v_{\mathrm{ind},0} \right),
\end{equation}
where $\rho$ is air density, $A = \pi R^2$ is rotor disk area, and $V_\infty$ is the freestream velocity. This provides a global constraint linking thrust and induced velocity but does not resolve blade-level forces.

From blade element theory, each blade of chord $c$ is discretized into radial segments of length $\mathrm{d}r$, and the elemental thrust and torque are:
\begin{align}
\mathrm{d}T &= \frac{1}{2} \rho U_{\mathrm{rel}}^2(r) \, c \, C_L(\alpha) \, \mathrm{d}r, \\
\mathrm{d}Q &= \frac{1}{2} \rho U_{\mathrm{rel}}^2(r) \, c \, C_D(\alpha) \, r \, \mathrm{d}r,
\end{align}
where $U_{\mathrm{rel}}(r)$ is the local relative velocity (combination of rotational and inflow components), $\alpha$ is the local angle of attack, and $C_L(\alpha)$, $C_D(\alpha)$ are airfoil lift and drag coefficients.

BEMT couples the two: momentum theory provides $v_{\mathrm{ind}}(r)$, the induced velocity distribution, while blade element theory provides $C_L$, $C_D$ from airfoil data and calculates the elemental forces. An iterative procedure ensures that the thrust computed from the integrated blade element loads matches the momentum-theory prediction. This model captures both induced and profile power losses and allows for the inclusion of rotor-specific geometric parameters.

\section{Dryden Response model}\label{appx:dryd}
The Dryden model provides continuous-time shaping filters that, when driven by Gaussian white noise, generate turbulence velocity components $(u_{\mathrm{turb}}, v_{\mathrm{turb}}, w_{\mathrm{turb}})$ with statistical properties matching the Dryden power spectral densities (PSDs). The one-sided PSDs for longitudinal $(u_{\mathrm{turb}})$, lateral $(v_{\mathrm{turb}})$, and vertical $(w_{\mathrm{turb}})$ turbulence components are:
\begin{align}
\Phi_u(\Omega_{\mathrm{sp}}) &= \sigma_u^2 \frac{2 L_u / \pi}{1 + (L_u \Omega_{\mathrm{sp}})^2}, \\
\Phi_v(\Omega_{\mathrm{sp}}) &= \sigma_v^2 \frac{L_v / \pi}{1 + (L_v \Omega_{\mathrm{sp}})^2} 
                 \left( 1 + \frac{4 L_v^2 \Omega_{\mathrm{sp}}^2}{1 + (L_v \Omega_{\mathrm{sp}})^2} \right), \\
\Phi_w(\Omega_{\mathrm{sp}}) &= \sigma_w^2 \frac{L_w / \pi}{1 + (L_w \Omega_{\mathrm{sp}})^2} 
                 \left( 1 + \frac{3 L_w^2 \Omega_{\mathrm{sp}}^2}{1 + (L_w \Omega_{\mathrm{sp}})^2} \right),
\end{align}
where $\sigma_{\{\cdot\}}$ are turbulence standard deviations, $L_{\{\cdot\}}$ are turbulence scale lengths, and $\Omega_{\mathrm{sp}}$ is the spatial frequency in the aircraft frame.  

From these PSDs, transfer functions generate wind velocity fluctuations from white-noise inputs. The turbulence components are then expressed in the inertial frame and transformed into the rotor disk frame.

\section{Runge-Kutta Scheme}\label{appx:rks}
Given $\dot{\mathbf{x}} = \mathbf{f}(t,\mathbf{x})$ (collecting translational
and rotational dynamics) and step $h$:
\begin{align}
\mathbf{k}_1 &= \mathbf{f}(t_n,\mathbf{x}_n), \\
\mathbf{k}_2 &= \mathbf{f}\!\left(t_n+\tfrac{h}{2},\,\mathbf{x}_n+\tfrac{h}{2}\mathbf{k}_1\right), \\
\mathbf{k}_3 &= \mathbf{f}\!\left(t_n+\tfrac{h}{2},\,\mathbf{x}_n+\tfrac{h}{2}\mathbf{k}_2\right), \\
\mathbf{k}_4 &= \mathbf{f}\!\left(t_n+h,\,\mathbf{x}_n+h\,\mathbf{k}_3\right), \\
\mathbf{x}_{n+1} &= \mathbf{x}_n + \tfrac{h}{6}\left(\mathbf{k}_1 + 2\mathbf{k}_2 + 2\mathbf{k}_3 + \mathbf{k}_4\right).
\end{align}

\section{Propagation model}\label{appx:atm_direct}
Atmospheric absorption is modeled according to ISO 9613-1:1993 \cite{ISO9613_1_1993}, which expresses the attenuation per unit distance as a frequency-dependent coefficient $\alpha(f)$ in~dB/m.  
The per-band attenuation over a propagation distance $d$ is given by:
\begin{equation}
    A_{\mathrm{atm}}(f, d) = \alpha(f) \cdot d
\end{equation}

The absorption coefficient $\alpha(f)$ is computed as:
\begin{equation}
\begin{split}
    \alpha(f) &= 8.686\, f^2 \Bigg[
    1.84 \times 10^{-11} \frac{1}{p_r} \left( \frac{T}{T_0} \right)^{1/2} \\
    &\quad + \left( \frac{T}{T_0} \right)^{-5/2}
    \left( \frac{0.01275 \, e^{-2239.1/T}}{f_{\mathrm{rO}} + \frac{f^2}{f_{\mathrm{rO}}}}
    + \frac{0.1068 \, e^{-3352/T}}{f_{\mathrm{rN}} + \frac{f^2}{f_{\mathrm{rN}}}} \right)
    \Bigg]
\end{split}
\end{equation}

Here, $f$ is the center frequency of the 1/3-octave band measured in [Hz], $p_r$ is the relative atmospheric pressure with respect to $101.325\,\text{[kPa]}$, $T$ is the absolute temperature in [K] with $T_0 = 293.15\,\text{K}$, and $f_{\mathrm{rO}}$ and $f_{\mathrm{rN}}$ are the oxygen and nitrogen relaxation frequencies.

\begin{equation}
    f_{\mathrm{rO}} = p_r \left[ 24.0 + \frac{4.04 \times 10^4 \, h \, (0.02 + h)}{0.391 + h} \right]
\end{equation}
\begin{equation}
    f_{\mathrm{rN}} = p_r \left( \frac{T}{T_0} \right)^{-1/2}
    \left[ 9.0 + 280.0\, h\, e^{-4.170 \left( (T/T_0)^{-1/3} - 1 \right)} \right]
\end{equation}

The molar humidity ratio $h$ is given by:
\begin{equation}
    h = \frac{H_r \, P_{\mathrm{sat}}(T)}{p_{\mathrm{atm}}}
\end{equation}
where $H_r$ is the relative humidity (0--1), $p_{\mathrm{atm}}$ is the ambient pressure in~kPa, and $P_{\mathrm{sat}}(T)$ is the saturation vapor pressure at temperature $T$.

When source directivity is not already embedded in the reference sound power data, a Directivity Index (DI) correction term can be applied per band. This correction accounts for directional radiation characteristics as a function of observation angle $\theta$ and frequency $f$.

The DI is defined as:
\begin{equation}
    DI(f, \theta) = 10 \log_{10} \left[ \frac{I(f, \theta)}{\overline{I}(f)} \right]
\end{equation}
where $I(f, \theta)$ is the radiated acoustic intensity in the direction $\theta$, and $\overline{I}(f)$ is the average intensity over the full sphere:
\begin{equation}
    \overline{I}(f) = \frac{1}{4\pi} \int_{0}^{2\pi} \int_{0}^{\pi} I(f, \theta) \, \sin\theta \, d\theta \, d\phi
\end{equation}

The directivity term is added to the per-band propagation equation for the sound pressure level at the receiver:
\begin{equation}
    L_p(f) = L_w(f) - A_{\mathrm{sp}}(d) - A_{\mathrm{atm}}(f, d) + DI(f, \theta)
\end{equation}
where $A_{\mathrm{sp}}(d)$ is the free-field spherical spreading attenuation, $A_{\mathrm{atm}}(f, d)$ is the atmospheric absorption, and $DI(f, \theta)$ is expressed in decibels.

\section{Simulation parameters}\label{appx:param}

The simulation setup is defined by a set of parameters that specify the temporal resolution, mission duration, vehicle dynamics, and environmental constraints. Table~\ref{tab:sim_params} summarizes all the values adopted in this study.

The simulation time step is set to \(dt = 0.008\)~s, with a total mission duration of 150~s. Thresholds for trajectory switching and simulation termination are set to 2~m. A circular region of 14~m radius is considered for evaluating noise annoyance.

The quadrotor model assumes four rotors, each with a maximum speed of 3000~RPM. Maximum roll, pitch, and yaw angles are constrained to \(30^\circ\). Velocity limits are set to 50~km/h horizontally and 20~km/h vertically. Yaw target is fixed to $0$. The vehicle mass is 5.2~kg, with inertia and aerodynamic coefficients listed in Table~\ref{tab:sim_params}.

Noise predictions rely on pre-trained surrogate models, with the rotor speed normalized to a reference of 2500~RPM. 

Finally, several optimization algorithms are employed for controller tuning. GA, PSO, and GWO are configured with common settings such as population size, number of iterations, and rates for crossover, mutation, and elitism. Their specific configurations are also summarized in Table~\ref{tab:sim_params}.

\begin{table}[H]
    \centering
    \caption{Simulation parameters used in the experiments.}
    \label{tab:sim_params}
    \begin{tabular}{ll}
        \toprule
        Parameter & Value \\
        \midrule
        Time step $dt$ & 0.008~s \\
        Simulation time & 150~s \\
        Termination threshold & 2~m \\
        Target shift threshold & 2~m \\
        Noise annoyance radius & 14~m \\
        Max RPM & 3000 \\
        Number of rotors & 4 \\
        Number of blade per rotor & 2 \\
        Rotor diameter & 0.6m \\
        Rotor radius hub & 0.02m \\
        BEMT number of sections & 8 \\
        Max roll/pitch/yaw angle & $30^\circ$ \\
        Max horizontal speed & 50~km/h \\
        Max vertical speed & 20~km/h \\
        Mass & 5.2~kg \\
        Inertia & [3.8e-3, 3.8e-3, 7.1e-3]~kg\,m$^2$ \\
        Arm length $l$ & 0.32~m \\
        Rotor drag coefficient $d$ & $7.5\times 10^{-7}$ \\
        Traslational drag coefficients. $C_d$ & [0.1, 0.1, 0.15] \\
        Aerodynamic friction coefficients. $C_a$ & [0.1, 0.1, 0.15] \\
        Rotor inertia $J_r$ & $6\times 10^{-5}$~kg\,m$^2$ \\
        Reference RPM (noise) & 2500 \\
        GA settings & 30 pop., $c_r=0.8$, $m_r=0.1$, elitism 0.1 \\
        PSO settings & 30 swarm, $w=0.7$, $c_1=1.5$, $c_2=1.5$ \\
        GWO settings & 30 wolves \\
        \bottomrule
    \end{tabular}
\end{table}

\bibliographystyle{elsarticle-num-names} 
\bibliography{biblio_2}

\end{document}